\documentclass[12pt]{article}
\def\Red{} % PANTONE RED
\def\Blue{} % PANTONE BLUE-072
\def\Black{} % PANTONE PROCESS-BLACK
\usepackage{amssymb}
\usepackage{amsfonts}
\usepackage{mathrsfs}
\usepackage{latexsym}
\usepackage{amsmath}
\usepackage{graphics}
\usepackage{graphicx}
\usepackage{sectsty}
\usepackage{setspace}
\usepackage{multirow}
\sectionfont{\normalsize\bf}
\subsectionfont{\rm\normalsize}
\def\be{\begin{equation}}
\def\ee{\end{equation}}
\input amssym.def
\input amssym
\def\U{{\cal U}}
\def\V{{\cal V}}
\def\W{{\cal W}}
\def\R{{\cal R}}
\def\A{{\cal A}}
\def\O{{\cal O}}
\def\P{{\cal P}}
\def\C{{\cal C}}
\def\M{{\cal M}}
\def\B{{\cal B}}

\def\N{{\cal N}}
\def\T{{\cal T}}
\def\Sc{{\cal S}}
\def\bpi{{\overline\pi}}

\def\vss{\vskip12pt}
\def\bS{{\overline{\cal S}}}
\def\bT{{\overline{\cal T}}}

\def\fup{\vbox{\vskip-1pt\hbox{\Red$\otimes$\Black}\vskip1pt} }
\def\fdown{\vbox{\vskip-1pt\hbox{\Blue$\boxtimes$\Black}\vskip1pt}  }

\begin{document}

\thispagestyle{empty}
\begin{flushright}
\end{flushright}
\baselineskip=16pt
\vspace{.5in}
{%\Large
\begin{center}
{\bf General Split Helicity Gluon Tree Amplitudes in Open Twistor String 
Theory}
\end{center}}
\vskip 1.1cm
\begin{center}
{Louise Dolan}
\vskip5pt

\centerline{\em Department of Physics}
\centerline{\em University of North Carolina, Chapel Hill, NC 27599} 
\bigskip
\bigskip        
{Peter Goddard}
\vskip5pt

\centerline{\em Institute for Advanced Study}
\centerline{\em Princeton, NJ 08540, USA}
\bigskip
\bigskip
\bigskip
\bigskip
\end{center}

\abstract{\noindent 
We evaluate all split helicity gluon tree amplitudes in open twistor string theory. We show 
that these amplitudes satisfy the BCFW recurrence relations restricted to the split helicity case and, hence, that these amplitudes agree with those of gauge theory. To do this we make a particular choice of the sextic constraints in the link variables that determine the poles contributing to the contour integral expression for the amplitudes. Using the residue theorem to re-express  this integral in terms of contributions from poles at rational values of the link variables, which we  determine, we evaluate 
the  amplitudes explicitly, regaining the gauge theory results of Britto {\it et al.} \cite{Britto}.

\bigskip

%\vskip120pt email: ldolan@physics.unc.edu
\setlength{\parindent}{0pt}
\setlength{\parskip}{6pt}

\setstretch{1.05}
\vfill\eject
\vskip50pt
%%%%%%%%%%%%%%%%%%%%%%%%%%%%%%%%PAPER%%%%%%%%%%%%%%%
\section{\bf Introduction}

In this paper, we extend the techniques introduced in \cite{DG3} to evaluate explicitly the general split helicity gluon tree amplitudes in open twistor string theory, establishing the gauge theory recursion relations
\cite{BCF}-\cite{BCFW} for these twistor string amplitudes,. This approach is based on the use of the link variables of Arkani-Hamed, Cachazo, Cheung and Kaplan \cite{ACCK1,ACCK2}; see also \cite{ACC}-\cite{ABCT2}. 

Twistor string theory \cite{W}-\cite{BW} provides tree level four-dimensional
$N=4$ Yang-Mills theory with a potential  string description. 
Particular examples of the derivation of
gauge tree amplitudes from the twistor string are given in \cite{BerkMotl} -\cite{N},
and recent ones using link variables are given in \cite{DG3}, \cite{SV} and \cite{NVW}. 
Gauge field theory formulae for split helicity trees are found in
\cite{Britto}-\cite{BDDK}.

We consider the twistor string tree amplitude with $m$ positive helicity gluons, labeled $i_1,\ldots,i_m$, and $n$ negative helicity gluons, labeled $r_1,\ldots, r_n$, with the helicities of the same sign being adjacent, {\it i.e.} the split helicity tree amplitude. Write $\P=\{i_1,\ldots,i_m\}$ and $\N=\{r_1,\ldots, r_n\}$; and the gluon momenta ${p^a_\alpha}_{\dot a}
=\pi_\alpha^a\bpi_{\alpha\dot a}.$ The link variables $c_{lu},l \in\P, u\in\N$ satisfy the $2(m+n)$ linear equations 
\begin{align}
\pi_l &=\sum_{u\in\N} c_{lu}\pi_u\label{L1}\\
\bar\pi_u &=-\sum_{l\in\P}\bar\pi_l c_{lu}.\label{L2}
\end{align}
where we have suppressed the spinor indices.
(See \cite{DG1, DG3} for our conventions.)
These equations are not independent because
they imply momentum conservation \cite{ACCK1}, and
for momenta satisfying this consistency condition they provide $2(m+n)-4$ 
constraints on the $mn$ variables $c_{lu}$, leaving 
$N_R=(m-2)(n-2)$ degrees of freedom. Fixing $i,j\in\P$ and $r,s\in\N$, we can take the independent degrees of freedom to be $c_{kt}$, where $k\in\P', t\in\N'$, where $\P'=\{k\in\P,k\ne i,j\}$ and
$\N'=\{t\in\N,t\ne r,s\}$, the remaining $c_{lu}$ being expressed in terms of these $N_R$ variables using (\ref{L1}, \ref{L2}).

In \cite{DG3} we showed how to write the general twistor string tree amplitude as a contour integral over the $N_R$ variables $c_{kt}$, $k\in\P', t\in\N'$,
\be \M_{mn}=K_{mn}\oint_\O F(c) \prod_{k\in\P'\atop t\in\N'} {dc_{kt}\over \C_{kt}},\label{ointF}\ee
where $K_{mn}=\langle r,s\rangle^{2-m}[i,j]^{2-n}$ and $F(c)$ is a simple rational function of the $c_{lu}$, the form of which is given by (4.12) of \cite{DG3},  the sextic functions $\C_{kt}\equiv\C^{ijk}_{rst}$ are given by the determinants, 
\begin{align}
\C^{ijk}_{rst}&=\left|\begin{matrix} c_{is}c_{it}&c_{it}c_{ir}& c_{ir}c_{is}\cr
c_{js}c_{jt}&c_{jt}c_{jr}& c_{jr}c_{js}\cr
c_{ks}c_{kt}&c_{kt}c_{kr}& c_{kr}c_{ks}\cr\end{matrix}\right|,
\label{C'}\end{align}
and the contour $\O$ is chosen so as to include the residue contributions
from each of the simultaneous zeros of the $\C_{kt}$,
but none of those arising from poles of $F(c)$. 
The $N_R$ conditions $\C_{kt}=0, k\in\P', t\in\N'$, which can be viewed as constraints on the $mn$ variables $c_{lu}$, are equivalent to the condition that the $m\times n$ matrix with entries $c^{-1}_{lu}, l\in\P, u\in\N$, has rank 2, which in turn is the condition that the link variables $c_{lu}$ are of the form implied by twistor string theory \cite{DG3}. 
In the split helicity case, $F(c)$ is given by (4.11) of \cite{DG3} when the fixed labels $i=i_1, j=i_m, r=r_1, s=r_n$.

The key observation in evaluating the split helicity amplitudes is that, in this case, the integrand of (\ref{ointF}) simplifies if we replace the $N_R$ constraints $\C_{kt}=\C^{ijk}_{rst}$, $k\in\P', t\in\N'$, by another independent set in which the labels are contiguous, namely, 
\be C_{ab}=\C^{i_{a-1}i_ai_{a+1}}_{r_{b-1}r_br_{b+1}},
\qquad 2\leq a\leq m-1,  \quad 2\leq b\leq n-1, \label{defC}\ee
where $\C^{i_{a-1}i_ai_{a+1}}_{r_{b-1}r_br_{b+1}}$ is defined by an equation of the form (\ref{C'}).
Using these contiguous constraints, 
\be \M_{mn}=K_{mn}\oint_\O \hat F(c) \prod_{a=2}^{m-1} \prod_{b=2}^{n-1} {dc_{ab}\over C_{ab}},\label{ointhF}\ee
where $c_{ab}=c_{i_ar_b}$, the contour $\O$ is now chosen so as to include the residue contributions
from each of the simultaneous zeros of the $C_{ab}$,
but none of those arising from poles of $\hat F(c)$, which has the simple form
\be \hat F(c)={\Psi(c)\over c^{i_1i_2}_{r_1r_2}c^{i_{m-1}i_{m}}_{r_{n-1}r_n}},\label{hatF}\ee
where $c^{kl}_{tu}=c_{kt}c_{lu}-c_{ku}c_{lt}$, and $\Psi(c)$ is a multinomial expression in the $c_{ab}$ that we define in section \ref{ContConst}. It is this simple form, with only two factors in the denominator, that makes the evaluation relatively straightforward.

In section 2,
we first calculate the factor $J(c)$ that relates the function $F(c)$, in the integrand of the amplitude, appropriate to our original choice of constraints \cite{DG3}, to the simpler integrand function $\hat F(c)$, appropriate to the choice of contiguous constraints (\ref{defC}); and then, second, we introduce a parametrization, $\hat\beta{ab}$ of the solutions of the linear conditions (\ref{L1}), (\ref{L2}) appropriate to the contiguous constraints and calculate the Jacobian necessary to write  $\M$ as an integral over $\hat\beta_{ab}$, 
\begin{align}
 \M_{mn}
 &=\hat K_{mn}\oint_\O \hat F(c) \prod_{a=2}^{m-1} \prod_{b=2}^{n-1} {d\hat\beta_{ab}\over C_{ab}},\label{ointhFc}\\
\hat K_{mn}&=\prod_{a=2}^{m-2} [i_a,i_{a+1}]^{n-2}
\prod_{b=2}^{n-2} \langle r_b,r_{b+1}\rangle^{m-2}. \label{Khat}\end{align}
In section 3, we show how the simple form of $\hat F(c)$ for the general split-helicity tree enables the amplitude to be written in terms of similar expressions involving fewer integrations using the global residue theorem. In this way, we inductively express $\M_{mn}$ in terms of a sum of terms, corresponding to Young diagrams, given by the residue of the integrand at poles specified by $N_R$ conditions of the form $c^{i_ai{a+1}}_{r_br_{b+1}}=0$. In section 4, we describe how to solve these conditions iteratively to give expressions for $c_{lu}$ at the poles. 

Using this analysis, we establish the BCFW recursion relation for these amplitudes in section 5, thus demonstrating that the twistor string theory yields the gauge theory tree amplitudes in the split helicity case. In section 6, we apply the results of section 4 to derive explicit expressions for the terms contributing to $\M_{mn}$, obtaining the gauge theory results of Britto {\it et al.} \cite{Britto}.
In section 7, we use our general expression to evaluate the particular cases of the (4,4) and  (5,3) split helicity trees. Section 8 contains some comments about the extension of our approach to non-split helicity tree amplitudes. 

\Black

\vss
\section{\bf Contiguous Constraints}
\label{ContConst}

\subsection{\sl The Integrand $\hat F(c)$}
\label{ContConst1}

For split helicity amplitudes, the integrand  $F(c)$ in (\ref{ointF}) is given by (4.11) in \cite{DG3},

\be
F(c) = {\left[c^{ij}_{rs}\right]^{N_R+1}\over  c_{ir} c_{js}}\prod_{a=1}^{m-1}
{1\over c^{i_ai_{a+1}}_{rs}}
\prod_{b=1}^{n-1}{1\over c^{ij}_{r_br_{b+1}}}\prod_{a=2}^{m-1}\prod_{b=2}^{n-1}
{1\over c_{i_ar_b}}
\prod_{a=2}^{m-1}c_{i_ar}^{n-2}c_{i_as}^{n-2}\prod_{b=2}^{n-1}c_{ir_b}^{m-2}
c_{jr_b}^{m-2},\label{fofc}\ee
where  $i_1 = i, \,
i_m = j,\,  r_1 =r, \, r_n =s$.

Now, since 
$$\C^{ijk}_{rst}= c^{ij}_{rs}c^{ik}_{st}c_{kr}c_{jt}-c^{ij}_{st}c^{ik}_{rs}c_{kt}c_{jr},$$
$$\C^{ijg}_{rst}=\C^{igk}_{rst}
{c^{ij}_{rs}c_{jt}\over c^{ik}_{rs}c_{kt}},\qquad\hbox{when}\quad\C^{ijk}_{rst}=0,$$
and, so for the purpose of calculating residues
at the zeros of constraints, we have the equivalences
 $$\C^{ijg}_{rst}\C^{ijk}_{rst} \sim \C^{igk}_{rst}\C^{ijk}_{rst}
{c^{ij}_{rs}c_{jt}\over c^{ik}_{rs}c_{kt}}
\sim \C^{igk}_{rst}\C^{jgk}_{rst}{c^{ij}_{rs}c_{it}c_{jt}
\over c^{kg}_{rs}c_{kt}c_{gt}}.$$

In this sense the product of constraints
\be\prod_{k\in \P'\atop t\in \N'} \C_{kt}
= \prod_{b=1}^{n-2}\prod_{a=1}^{m-2}\C^{i_1i_{a+1}i_m}_{r_1r_{b+1}r_n}\sim
J(c) \prod_{a=1}^{m-2}\prod_{b=1}^{n-2}\C^{i_{a}i_{a+1}i_{a+2}}_{r_{b}r_{b+1}
r_{b+2}},\ee
where
\begin{align}
J(c)&=\prod_{b=1}^{n-2}\prod_{a=2}^{m-2}{c^{ij}_{rs}c_{ir_{b+1}}c_{jr_{b+1}}
\over c^{i_{a}i_{a+1}}_{rs}c_{i_ar_{b+1}}c_{i_{a+1}r_{b+1}}}
\prod_{a=1}^{m-2}\prod_{b=2}^{n-2}{c^{i_{a}i_{a+1}}_{rs}c_{i_{a+2}r}
c_{i_{a+2}s}\over c^{i_{a}i_{a+1}}_{r_{b}r_{b+1}}c_{i_{a+2}r_{b}}c_{i_{a+2}
r_{b+1}}}\cr
&=\prod_{b=1}^{n-2}{c^{ii_{2}}_{rs}c_{i_2r_{b+1}} \over 
c^{ij}_{rs}c_{jr_{b+1}}}
\prod_{a=1}^{m-2}{c^{i_{a}i_{a+1}}_{rr_{2}}c_{i_{a+2}r_{2}} \over 
c^{i_{a}i_{a+1}}_{rs}c_{i_{a+2}s}}
\prod_{b=1}^{n-2}\prod_{a=1}^{m-2}{c^{ij}_{rs}c_{ir_{b+1}}c_{jr_{b+1}}
c_{i_{a+2}r}c_{i_{a+2}s}\over c^{i_{a}i_{a+1}}_{r_{b}r_{b+1}}c_{i_ar_{b+1}}
c_{i_{a+1}r_{b+1}}c_{i_{a+2}r_{b}}c_{i_{a+2}r_{b+1}}}.
\nonumber\end{align}
Using $${c^{ii_{2}}_{rs}c_{i_2r_{b+1}} \over c^{ij}_{rs}c_{jr_{b+1}}}
\sim{c^{ii_{2}}_{r_{b}r_{b+1}} c_{i_2r}c_{i_2s} c_{jr_b}\over c^{ij}_{r_{b}
r_{b+1}}c_{jr}c_{js}c_{i_2r_b}},$$
we have
\begin{align}
J(c) &\sim c^{ii_{2}}_{rr_2}c^{i_{m-1}j}_{r_{n-1}s}
{\left[c^{ij}_{rs}\right]^{N_R+1}\over  c_{ir}c_{js}}\Black
\prod_{b=1}^{n-1}{1 \over c^{ij}_{r_{b}r_{b+1}}}
\prod_{a=1}^{m-1}{1\over c^{i_{a}i_{a+1}}_{rs}}
\prod_{a=2}^{m-1}\left[c_{i_ar}c_{i_as}\right]^{n-2}\prod_{b=2}^{n-1}
\left[c_{ir_b}c_{jr_b}\right]^{m-2}\Black\cr
&\qquad\times\prod_{b=2}^{n-2}\prod_{a=2}^{m-2}{1\over 
c^{i_{a}i_{a+1}}_{r_{b}r_{b+1}}c_{i_ar_{b}}c_{i_{a+1}r_{b+1}}}
\prod_{b=2}^{n-2}{1\over c_{ir_{b}}c_{jr_{b+1}}}\prod_{a=2}^{m-2}{1\over 
c_{i_ar}c_{i_{a+1}s}}
\prod_{b=2}^{n-1}\prod_{a=2}^{m-1}{1\over c_{i_{a}r_{b}}}\Black   
\prod_{b=2}^{n-1}\prod_{a=2}^{m-1}{1\over c_{i_{a}r_{b}}}, 
\nonumber\end{align}

\be F(c)\left/\prod_{b=1}^{n-2}\prod_{a=1}^{m-2}\C^{i_1i_{a+1}i_m}_{r_1r_{b+1}r_n}
\right.
\sim \hat F(c)\left/\prod_{a=1}^{m-2}\prod_{b=1}^{n-2}\C^{i_{a}i_{a+1}
i_{a+2}}_{r_{b}r_{b+1}r_{b+2}}\right.,\label {FC}\ee

where $\displaystyle\hat F(c)={ F(c)\over J(c)}$, and
so that the relevant function for contiguous constraints is
\be\hat F(c)={1\over c^{i_1i_{2}}_{r_1r_2}c^{i_{m-1}i_m}_{r_{n-1}r_n}}
\prod_{b=2}^{n-2}\prod_{a=2}^{m-2}c^{i_{a}i_{a+1}}_{r_{b}r_{b+1}}c_{i_ar_{b}}
c_{i_{a+1}r_{b+1}}\prod_{b=2}^{n-2} c_{ir_{b}}c_{jr_{b+1}}\prod_{a=2}^{m-2} 
c_{i_ar}c_{i_{a+1}s}
\prod_{b=2}^{n-1}\prod_{a=2}^{m-1} c_{i_{a}r_{b}}.\label{Fhat}\ee
Writing
$ f_{ab}= c^{i_ai_{a+1}}_{r_br_{b+1}} $,
the integrand becomes 
\begin{align}
 I_{mn} &= {\hat F(c)\over  \prod_{b=2}^{n-1}\prod_{a=2}^{m-1}
\C^{i_{a-1}i_ai_{a+1}}_{r_{b-1}r_br_{b+1}}}\cr
&={1\over f_{11}f_{m-1,n-1}}
\prod_{b=2}^{n-2}\prod_{a=2}^{m-2}f_{ab}c_{ab}c_{a+1,b+1}
\prod_{b=2}^{n-2} c_{1b}c_{m,b+1}\prod_{a=2}^{m-2} c_{a1}c_{a+1,n}
\prod_{b=2}^{n-1}\prod_{a=2}^{m-1} {c_{ab}\over C_{ab}},
\label{bigI}\end{align}
and for MHV amplitudes, this gives
\be I_{m2} = {1\over c_{11} c_{m2}} \prod_{a=1}^{m-1}
{1\over f_{a1}},\qquad
I_{2n} = {1\over c_{11} c_{2n}} \prod_{b=1}^{n-1}
{1\over f_{1b}}.\label{mhv}\ee

For low values of $m, n$, we have the following:
\be  I_{33}={c_{22}\over f_{11}f_{22}C_{22}};\quad I_{43}={c_{21}c_{22}c_{32}c_{33}\over f_{11}f_{32}C_{22}C_{32}};\label{I3343}\ee
\be I_{53}={c_{21}c_{22}c_{31}c_{32}c_{33}c_{42}c_{43}\over  f_{11}f_{42}C_{22}C_{32}C_{42}};\quad
I_{44}={f_{22}c_{12}c_{21}c_{22}^2c_{23}c_{32}c_{33}^2c_{34}c_{43}\over 
f_{11}f_{33}C_{22}C_{23}C_{32}C_{33}}.\label{I5344}\ee

\subsection{\sl  Parameterization of $c_{lu}$}
\label{ContConst2}

In \cite{DG3} it was shown that the general solution for $c_{lu}$ of the linear conditions (\ref{L1}), (\ref{L2}) can be written in the form 
\be c_{lu}= a_{lu}+{1\over 4 }\sum_{g,h\in\P\atop v,w\in\N}
\beta^{lgh}_{uvw}[g,h]\langle v,w\rangle, \label{beta}\ee
where $a_{lu}$ is a particular solution and $\beta^{lgh}_{uvw}$ is antisymmetric under permutations of $l, g, h$ and also under permutations of $u, v, w$. (Note this definition of 
$\beta^{lgh}_{uvw}$ differs by a factor of $p^2$ from that used in \cite{DG3}.) Because there are only $N_R$ independent solutions to (\ref{L1}), (\ref{L2}),
there is some arbitrariness in the choice of the parameters $\beta^{lgh}_{uvw}$; in \cite{DG3} this arbitrariness was resolved by requiring that 
$\beta^{lgh}_{uvw}=0$ unless $i,j\in\{l,g,h\}$ and $r,s\in\{u,v,w\}$, so that the all the $\beta^{lgh}_{uvw}$ are related to the $N_R$ parameters
$$\beta_{ab}=\beta^{i i_a j}_{r r_b s},\qquad 2\leq a\leq m-1, \quad 2\leq b\leq n-1,$$
with $i=i_1,j=i_m,r=r_1,s=r_n.$

This choice of parameters is appropriate when using the constraints $\C_{kt}$, but, when using contiguous constraints $C_{ab}$, it is more appropriate to replace  $\beta^{lgh}_{uvw}$ with parameters $\hat\beta^{lgh}_{uvw}$ such that $\hat\beta^{lgh}_{uvw}=0$ unless
$ \{l,g,h\}=\{i_{a-1},i_a,i_{a+1}\}$, for some $a$, where $2\leq a\leq m-1$, 
and $\{u,v,w\}=\{r_{b-1},r_b,r_{b+1}\}$, for some $b$, where $2\leq b\leq n-1$. Write 
$$\hat\beta_{ab}=\beta^{i_{a-1}i_a i_{a+1}}_{r_{b-1} r_b r_{b+1}},\qquad
2\leq a\leq m-1, \quad 2\leq b\leq n-1;$$
then, for $2\leq a\leq m-1$  and $2\leq b\leq n-1$,
\begin{align}
 c_{i_ar_b}- a_{i_ar_b}&=\beta_{ab}[i,j]\langle r,s\rangle\cr
 &=\sum_{a'=-1,0,1\atop 2\leq a+a'\leq m}
\sum_{b'=-1,0,1\atop 2\leq b+b'\leq n}(-1)^{a'+b'}\hat\beta_{a+a',b+b'}
[i_{\sigma(a,a')},i_{\tau(a,a')}]\langle r_{\sigma(b,b')},r_{\tau(b,b')}\rangle,\label{cbb}\cr
\end{align}
where
 $\sigma(a,-1)=a-2, \, \sigma(a,0)=a-1, \,\sigma(a,1)=a+1$ and 
$\tau(a,-1)=a-1, \,\tau(a,0)=a+1, \,\tau(a,1)=a+2$.

From (\ref{ointhF}),
\begin{align}
 \M_{mn}&=\tilde K_{mn}\oint_\O \hat F(c) \prod_{a=2}^{m-1} \prod_{b=2}^{n-1} {d\beta_{ab}\over C_{ab}},\qquad \tilde K_{mn}=[i,j]^{(m-3)(n-2)}
 \langle r,s\rangle^{(m-2)(n-3)},\cr
 &=\tilde K_{mn}\oint_\O \hat F(c) \left|{\partial\beta\over\partial\hat\beta}\right|\prod_{a=2}^{m-1} \prod_{b=2}^{n-1} {d\hat\beta_{ab}\over C_{ab}}.
 \label{ointhFb}\end{align}
The Jacobian can be computed in two stages. First define
$$\tilde\beta_{ab}\langle r,s\rangle=\sum_{b'=-1,0,1\atop 2\leq b+b'\leq n}
(-1)^{b'}\hat\beta_{a,b+b'}\langle r_{\sigma(b,b')},r_{\tau(b,b')}\rangle,$$
then
$$\beta_{ab}[i,j]=\sum_{a'=-1,0,1\atop 2\leq a+a'\leq m}(-1)^{a'}\tilde
\beta_{a+a',b}[i_{\sigma(a,a')},i_{\tau(a,a')}].$$
Now if $J_{m,b}[i,j]^{-m+2}$ denotes the Jacobian of $\beta_{ab}$ with respect 
to $\tilde\beta_{ab}$, $2\leq a\leq m-2$, $J_{m,b}$ satisfies the recurrence relation
\be
J_{m,b}=[i_{m-2},i_{m}]J_{m-1,b}-[i_{m-3},i_{m-2}][i_{m-1},i_{m}]J_{m-2,b},
\label{recur1}\ee
which
uniquely specifies $J_{m,b} $ subject to the conditions $J_{3,b}
=[i_1,i_3],$ $J_{4,b}=[i_1,i_4][i_2,i_3]$,
and the solution is provided by
$$J_{m,b}=[i_1,i_m]\prod_{a=2}^{m-2} [i_a,i_{a+1}].$$
Then, for fixed $b$, the $(m-2)\times (m-2)$ Jacobian 
$$\left|{\partial \beta\over\partial\tilde\beta}\right|
=\prod_{a=2}^{m-2} {[i_a,i_{a+1}]\over[i,j]} .$$
Similarly, for fixed $a$, the $(n-2)\times (n-2)$ Jacobian
$$\left|{\partial \tilde\beta\over\partial\hat\beta}\right|
=\prod_{b=2}^{n-2} {\langle r_b,r_{b+1}\rangle\over\langle r,s\rangle} ,$$
so that the full Jacobian 
\be\left|{\partial \beta\over\partial\hat\beta}\right|
=\left(\prod_{a=2}^{m-2} {[i_a,i_{a+1}]\over[i,j]}\right)^{n-2}
\left(\prod_{b=2}^{n-2} {\langle r_b,r_{b+1}\rangle\over\langle r,s\rangle}
\right)^{m-2},\label{jac1}\ee
leading to (\ref{ointhFc}) and (\ref{Khat}).

\section{\bf Towards a Recurrence Relation }
\label{RecRel}

Exploiting the simple form of the denominator of $\hat F(c)$, we can 
use the global residue theorem \cite{ACCK1, DG3} to 
reduce the number of integrations.  First we illustrate this for 
\be\M_{44} =\hat K_{44}
\oint_\O I_{44}  d\hat\beta_{22}d\hat\beta_{23}d\hat\beta_{32}d\hat\beta_{33},\label{amp4}\ee
where $I_{44}$ is given by (\ref{I5344}). To evaluate (\ref{amp4}) by the global residue theorem,
we divide the factors in the denominator of the integrand into disjoint subsets:
$$\Gamma_{22}=\{C_{22}\};\quad\Gamma_{23}=\{C_{23},f_{11},f_{33}\};\quad\Gamma_{32}=\{C_{32}\};\quad\Gamma_{33}=\{C_{33}\}, $$
\be \M_{4,4} = 
\R\begin{pmatrix} C_{22}&C_{23}\cr C_{32}&C_{33}\cr\end{pmatrix}
 = -\R\begin{pmatrix}
C_{22}&f_{11}\cr C_{32}&C_{33}\cr\end{pmatrix}
-\R\begin{pmatrix}C_{22}& f_{33}\cr C_{32}&C_{33}
\end{pmatrix},\label{res4}\ee
where we are using the matrix notation 
$$\R \begin{pmatrix} C_{22}&C_{23}\cr C_{32}&C_{33}\cr\end{pmatrix}$$
for the residue that would be denoted by $\R( C_{22},C_{23}, C_{32},C_{33})$ in the notation of 
\cite{DG3}.

Since $ C_{22}= f_{11} f_{22} c_{13} c_{31}- f_{12} f_{21} c_{11} c_{33}$,  then for  $ f_{11} =0$, $ C_{22} = -  f_{12} f_{21} c_{11} c_{33}$;
but $ c_{33}$ is in the numerator of the integrand and when $ f_{11}=0$, $ c_{11}c_{22}=  c_{12} c_{21}$, which is also in the numerator, so that,
when $ f_{11} =0$, the relevant zeros of $C_{22}$ occur at $f_{12}=0$ and at $f_{21}=0$. Thus
\be\R\begin{pmatrix}
  C_{22}& f_{11}\cr C_{32}& C_{33}\end{pmatrix}
=\R\begin{pmatrix} 
 f_{12}& f_{11}\cr C_{32}& C_{33}\end{pmatrix}
+ \R\begin{pmatrix}
 f_{21}& f_{11}\cr C_{32}& C_{33}\end{pmatrix}.
\label{Rff}\ee
Further, since $C_{32}= f_{21} f_{32} c_{23} c_{41}- f_{22} f_{31} c_{21} c_{43}$, then for $f_{21}=0$, 
$C_{32}= - f_{22} f_{31} c_{21} c_{43}$ and so the integrand for the last term in (\ref{Rff}) is, up to a constant
\be{c_{22}^3c_{23}c_{32}c_{33}c_{34}\over 
f_{11}f_{33}f_{12} f_{21}  C_{23} f_{31} c_{21} C_{33}},\label{ffint}\ee
so that we do not have to consider the zeros of $C_{32}$ corresponding to $f_{22}=0$ or $c_{43}=0$. The zero of $C_{32}$ corresponding to 
$c_{21}=0$ when $f_{11}=f_{21}=0$, entails $c_{11}c_{22}=c_{31}c_{22}=0$; the presence of $c_{22}$ in the numerator, means that we would need
$c_{11}=c_{21}=c_{31}=0$ for a nonzero residue. But, from (\ref{L2}), this would imply that $\bpi_{i_4}$ is parallel to $\bpi_{r_1}$, and so for generic momenta 
there is no contribution from $c_{21}=0$. 

This leaves us to consider the possibility of a contribution to the last term in (\ref{Rff}) from $f_{11}=f_{21}=f_{31}=0$. Then
\be c_{32}(\bpi_{r_2}c_{21}-\bpi_{r_1}c_{22})=c_{32}\bpi_{i_4}(c_{22}c_{41}-c_{21}c_{42})=0.\label{pipara}\ee
Since $c_{32}$ is in the numerator of (\ref{ffint}), it must be nonzero for a nonzero contribution and then (\ref{pipara}) implies that $\bpi_{r_1}$ and $\bpi_{r_2}$ are parallel. So the last term in (\ref{Rff}) vanishes, leaving
\be\R\begin{pmatrix}
  C_{22}& f_{11}\cr C_{32}& C_{33}\end{pmatrix}
=\R\begin{pmatrix} 
 f_{12}& f_{11}\cr C_{32}& C_{33}\end{pmatrix}.
\label{Rff2}\ee

Similarly
\be  \R\begin{pmatrix} 
 C_{22}& f_{33}\cr C_{32}& C_{33}\end{pmatrix}
=  \R\begin{pmatrix}
 C_{22}& f_{33}\cr C_{32}& f_{23}\end{pmatrix},\label{Rff3}\ee
so that (\ref{res4}) becomes
\be\M_{44} =  
\R\begin{pmatrix}
 C_{22}& C_{23}\cr C_{32}& C_{33}\end{pmatrix}
= \R\begin{pmatrix} f_{11}& f_{12}\cr C_{32}& C_{33}
\end{pmatrix}
+\R\begin{pmatrix}  C_{22}& f_{23}
\cr C_{32}& f_{33}\end{pmatrix},\label{res42}\ee
using the antisymmetry of the residue $\R$ on interchanging its arguments.
The integrands of the terms on the right hand side of (\ref{res42}) are:
\be\hbox{for}\quad \R\begin{pmatrix} f_{11}& f_{12}\cr C_{32}& C_{33}\end{pmatrix},\qquad
{ c_{21}  c_{22}^2 c_{23} c_{32} c_{33} c_{43}\over  f_{11} f_{12} c_{11}f_{13} f_{21} f_{33} C_{32} C_{33}}= {1 \over f_{11} f_{12}}I_{34} 
{ c_{21} c_{22} c_{23} \over f_{13}  c_{11}},\label{rec1}\ee
where $I_{34}$ by defined from (\ref{bigI}) with the $i_a$ shifted to $i_{a+1}$, and
\be\hbox{for}\quad \R\begin{pmatrix}
 C_{22}& f_{23}\cr C_{32}& f_{33}\end{pmatrix},
\qquad
{ c_{21} c_{22} c_{23}c_{32} c_{33}^2c_{43}\over f_{23} f_{33} c_{44}f_{11}  f_{13} f_{32}  C_{22} C_{32}}
= {1 \over  f_{23} f_{33}}I_{43} { c_{23} c_{33} c_{43}  \over  f_{13}  c_{44}}.\label{rec2}\ee 
We shall see in section \ref{Jacobians} that the factors $c_{21} c_{22} c_{23} /f_{13}  c_{11}$ and $c_{23} c_{33} c_{43} /f_{13}  c_{44}$ in (\ref{rec1}) and (\ref{rec2}), respectively, correspond to appropriate Jacobians, so that (\ref{res42}) expresses $\M_{44}$ in terms of $\M_{34}$ and $\M_{43}$.

In a similar fashion we can establish a relation for the general $(m,n)$ split helicity amplitude, 
\be \M_{m,n}  = 
\R\begin{pmatrix}
 C_{22}& C_{23}&\ldots& C_{2,n-2}& C_{2,n-1}\cr
 C_{32}& C_{33}&\ldots& C_{3,n-2}& C_{3,n-1}\cr
\vdots&\vdots&&\vdots&\vdots\cr
 C_{m-2,2}& C_{m-2,3}&\ldots& C_{m-2,n-2}& C_{m-2,n-1}\cr
 C_{m-1,2}& C_{m-1,3}&\ldots& C_{m-1,n-2}& C_{m-1,n-1}
\end{pmatrix},\label{bigM0}\ee
 whose integrand is given by  (\ref{bigI}).

As in (\ref{res4}), we have
\begin{align}
&\M_{m,n}  =  
- \R\begin{pmatrix}
 C_{22}& C_{23}&\ldots& C_{2,n-2}& f_{11}\cr
 C_{32}& C_{33}&\ldots& C_{3,n-2}& C_{3,n-1}\cr
\vdots&\vdots&&\vdots&\vdots\cr
 C_{m-2,2}& C_{m-2,3}&\ldots& C_{m-2,n-2}& C_{m-2,n-1}\cr
 C_{m-1,2}& C_{m-1,3}&\ldots& C_{m-1,n-2}& C_{m-1,n-1}\cr
\end{pmatrix}\cr
&\hskip30truemm- \R \begin{pmatrix}
 C_{22}& C_{23}&\ldots& C_{2,n-2}& f_{m-1,n-1}\cr
 C_{32}& C_{33}&\ldots& C_{3,n-2}& C_{3,n-1}\cr
\vdots&\vdots&&\vdots&\vdots\cr
 C_{m-2,2}& C_{m-2,3}&\ldots& C_{m-2,n-2}& C_{m-2,n-1}\cr
 C_{m-1,2}& C_{m-1,3}&\ldots& C_{m-1,n-2}& C_{m-1,n-1}\cr
\end{pmatrix}. \label{bigM}\end{align}
Suppose first that $m, n\geq 4$. 
When $ f_{11}=0$, $ C_{22}=- f_{12} f_{21} c_{11} c_{33}$, for the same reason as in the case $m=n=4$ just discussed, the relevant zeros of $C_{22}$ occur at 
$f_{12}=0$ and at $f_{21}=0$, so that the first term on the right hand side of (\ref{bigM}) can be written as the sum of two terms replacing $C_{22}$ by $f_{12}$ and $f_{21}$, respectively. In the case $f_{21}=0$, $C_{32}= - f_{22} f_{31} c_{21} c_{43}$; $f_{22}, c_{43}$ are in the numerator and, if $m>4$, $c_{21}$ is as well, and the only zero that needs to be considered is those of $f_{31}$. We can then proceed inductively: if $f_{a1}=0$, with $1\leq a\leq m-2$, then $C_{a+1,2}= - f_{a2} f_{a+1,1} c_{a1} c_{a+2,3}$, and $f_{a2},  c_{a+2,3}$ are in the numerator and $c_{a1}$ is as well if $1\leq a< m-2$, so that the relevant zero comes from $f_{a+1,1}=0$. When we eventually reach $a=m-2$ (as we do immediately if $m=4$), $c_{m-2,1}$ is not in the numerator; but if $c_{m-2,1}=0$ and $f_{b1}=0, 1\leq b\leq m-2$, we are led to $c_{1b}=0, 1\leq b\leq m-1$, as in the $m=n=4$ case, which is excluded because it would imply that $\bpi_{i_m}$ is parallel to $\bpi_{r_1}$. So for the first term on the  right hand side of (\ref{bigM}) to be nonzero we need $f_{b1}=0, 1\leq b\leq m-1$, but then a similar argument to that used in (\ref{pipara}) would imply that $\bpi_{r_1}$ and $\bpi_{r_2}$ are parallel, so excluding this possibility. 

It follows that, in the first term on the right hand side of (\ref{bigM}), we only need to consider the zeros of $C_{22}$ corresponding to $f_{12}=0$. In this case, arguments like those just applied to the constraint functions in the first column can now be applied along the first column, to deduce that the relevant zeros of $C_{2b}$ come from $f_{1b}=0, 2\leq b\leq n-2$. The difference is that the final constraint $C_{2,n-1}$ is absent, so in this instance we are {\it not} led to infer $\pi_{i_1}$ and $\pi_{i_2}$ are parallel, which would have excluded this contribution as well. Exactly similar arguments applied to the second term on the right hand side of (\ref{bigM}) lead to the conclusion that the corresponding contributions come from replacing $C_{a+1, n-1}$ by $f_{a,n-1}, 2\leq a\leq m-2$, so that  (\ref{bigM}) becomes 
\begin{align}
&\M_{mn}  = 
(-1)^n \R\begin{pmatrix}
 f_{11}& f_{12}&\ldots& f_{1,n-3}& f_{1,n-2}\cr
 C_{32}& C_{33}&\ldots& C_{3,n-2}& C_{3,n-1}\cr
\vdots&\vdots&&\vdots&\vdots\cr
 C_{m-2,2}& C_{m-2,3}&\ldots& C_{m-2,n-2}& C_{m-2,n-1}\cr
 C_{m-1,2}& C_{m-1,3}&\ldots& C_{m-1,n-2}& C_{m-1,n-1}\cr
\end{pmatrix}\cr
&\hskip26truemm+(-1)^m \R \begin{pmatrix}
 C_{22}& C_{23}&\ldots& C_{2,n-2}& f_{2,n-1}\cr
 C_{32}& C_{33}&\ldots& C_{3,n-2}& f_{3,n-1}\cr
\vdots&\vdots&&\vdots&\vdots\cr
 C_{m-2,2}& C_{m-2,3}&\ldots& C_{m-2,n-2}& f_{m-2,n-1}\cr
 C_{m-1,2}& C_{m-1,3}&\ldots& C_{m-1,n-2}& f_{m-1,n-1}\cr
\end{pmatrix}, \label{bigMM}\end{align}
where we have again used the antisymmetry of the residue. 

It is straightforward to establish this relation also for $m=3$, 
$$\M_{3n}  = 
(-1)^n  \R ( f_{11},f_{12},\ldots, f_{1,n-3}, f_{1,n-2})
- \R(C_{22},C_{23},\ldots,C_{2,n-2}, f_{2,n-1}),$$
using similar arguments, and likewise for $n=3$.

Generalizing (\ref{rec1}) and (\ref{rec2}), the two terms in the relation (\ref{bigMM}) for $\M_{mn}$ have integrands 
\be
(-1)^n\left[\prod_{b=1}^{n-2}  {1\over f_{1b}}\right]I_{m-1,n}{1\over c_{11} f_{1,n-1}}\prod_{b=1}^{n-1} c_{2b}\label{recc1}\ee
and
\be (-1)^m\left[\prod_{a=2}^{m-1}  {1\over f_{a,n-1}}\right]I_{m,n-1}{1\over c_{mn} f_{1,n-1}}\prod_{a=2}^{m} c_{a,n-1},\label{recc2}\ee
respectively. Again, we shall see in section \ref{Jacobians} that the final factors in (\ref{recc1}) and (\ref{recc2}) are appropriate Jacobians. 

By applying (\ref{bigMM}) iteratively we can express $\M_{mn}$ in terms of a sum of contributions from poles all specified by conditions of the form 
\be f_{ab}\equiv c_{ab}c_{a+1,b+1}- c_{a,b+1}c_{a+1,b}=0.\label{fab}\ee
The iterative process has to be continued with the various terms generated until each term involves $I_{m'n'}$ with either $m'=2$ or $n'=2$, at which point all the constraint functions $C_{ab}$ have been removed and each term must involve $N_R=(m-2)(n-2)$ conditions of the form (\ref{fab}).

The various terms are distinguished by the particular set of 
$N_R$ conditions  (\ref{fab}). Not all sets of $N_R$ conditions of this form are permitted. To characterize the particular set of conditions associated with a particular term in the expression for $\M_{mn}$, we start with a $(m-2)\times (n-2)$ matrix or grid, which we label by pairs $\U=\{(a,b): 2\leq a\leq m-1,\;2\leq b\leq n-1\}$. The first term in (\ref{bigMM}), {\it i.e.} (\ref{recc1}), corresponds to applying the conditions $f_{1,b-1}=0, \;2\leq b\leq n-1$, which we can associate with the top row of our grid, and second term in (\ref{bigMM}), {\it i.e.} (\ref{recc2}), corresponds to applying the conditions $f_{a,n-1}=0, 2\leq a\leq m-1$, which we can associate with the right hand column of our grid. We denote the first by placing the symbol \fup in each of the places or squares on the first row of the grid and the second by placing the symbol \fdown in each of the squares on the last column. 
As we iteratively apply the relation  (\ref{bigMM}), we fill either the top row or the last column of the grid formed by unfilled squares. In this way we obtain a diagram of the form shown in Figure 1. 

\setlength{\unitlength}{1.5mm}
\centerline{\begin{picture}(36,32)
\linethickness{0.075mm}
\multiput(0,0)(4,0){10}
{\line(0,1){28}}
\multiput(0,0)(0,4){8}
{\line(1,0){36}}
\multiput(1,25)(4,0){9}{\fup}
\multiput(1,21)(4,0){8}{\fup}
\multiput(1,17)(4,0){6}{\fup}
\multiput(1,13)(4,0){6}{\fup}
\multiput(1,9)(4,0){3}{\fup}
\multiput(1,5)(4,0){3}{\fup}
\multiput(1,1)(4,0){2}{\fup}
\multiput(33,21)(4,0){1}{\fdown}
\multiput(25,17)(4,0){3}{\fdown}
\multiput(25,13)(4,0){3}{\fdown}
\multiput(13,9)(4,0){6}{\fdown}
\multiput(13,5)(4,0){6}{\fdown}
\multiput(9,1)(4,0){7}{\fdown}
	\end{picture}}
	\vskip6pt
\centerline{Figure 1. \sl A Young-like Diagram}
\vskip6pt

Generically, in such a diagram the 
\fup may or may not stretch all the way across the top row, and may or may not stretch all the way down the first column, starting from the top left $(2,2)$. Similarly, the  \fdown may or may not stretch all the way up the last column, and may or may not stretch all the way back along the bottom row starting from $(m-1,n-1)$.
However, either the  \fup must stretch all the way across the top row or the  \fdown must stretch all the way up the last column.
More generally, if $\V$ denotes the subset of the grid $\U$ occupied by \fup, the constraint on the allowable sets of conditions is that if $(a,b)\in\V$ then $(c,d)\in\V$ whenever $c\leq a$ and $d\leq b$. Similarly, if $\W$ denotes the complementary subset of $\U$ to $\V$, {\it i.e.} the squares occupied by \fdown, then if $(a,b)\in\W$ and $a\leq c$ and $b\leq d$ then $(c,d)\in\W$. The set of conditions characterizing the term are then
\be f_{a-1,b-1}=0, \;(a,b)\in\V,\; \qquad f_{ab}=0, \;(a,b)\in\W.\label{VW}\ee 

The conditions on the grid ensure that it has the form of a set of steps, {\it i.e.} the form of  a Young diagram \cite{FH}. The number of such diagrams is 
\be{(m+n-4)!\over (m-2)!(n-2)!},\label{numt}\ee
 so that this is the number of terms in the expression for $\M_{mn}$. The diagrams obtained in this way are in correspondence with the zigzag diagrams introduced by Britto {\it et al.} \cite{Britto} for the gauge theory, as would be expected. 
 
In some ways it is preferable to replace the diagrams of the form of Figure 1 by equivalent ones drawn in an $m\times n$ grid: we draw a box linking the centers of the squares $(a,b), (a,b+1), (a+1,b), (a+1,b+1)$ if $f_{ab}=0$ is one of our conditions, which is equivalent to having $(a+1,b+1)\in\V$ or $(a,b)\in\W$. Then, corresponding to Figure 1, we have the alternative representation shown in Figure 2, where we have labelled the rows and the columns by the corresponding positive and negative helicities respectively. 

This {\it step diagram} has the property that if the squares $(a,b), (a,d), (c,b), (c,d)$ are linked by a box then $c^{ac}_{bd}=c_{ab}c_{cd}-c_{ad}c_{bc}=0$. [It is possible to have $c^{ac}_{bd}=0$ and $c^{ac}_{de}=0$ without having $c^{ac}_{be}=0$ because the first two equations will hold if $c_{ad}=c_{cd}=0$, but we exclude such configurations here because the factors in the numerator of our integrands mean that they are not relevant.]

\setlength{\unitlength}{1.5mm}
\centerline{\begin{picture}(40,42)
\linethickness{0.075mm}
\multiput(0,0)(4,0){12}{\line(0,1){36}}
\multiput(0,0)(0,4){10}{\line(1,0){44}}
\Red
\linethickness{0.3mm}
\multiput(2,30)(4,0){10}{\line(0,1){4}}
\multiput(2,30)(0,4){2}{\line(1,0){36}}
\multiput(2,26)(4,0){9}{\line(0,1){4}}
\multiput(2,26)(0,4){2}{\line(1,0){32}}
\multiput(2,18)(4,0){7}{\line(0,1){8}}
\multiput(2,18)(0,4){3}{\line(1,0){24}}
\multiput(2,10)(4,0){4}{\line(0,1){8}}
\multiput(2,10)(0,4){3}{\line(1,0){12}}
\multiput(2,6)(4,0){3}{\line(0,1){4}}
\multiput(2,6)(0,4){2}{\line(1,0){8}}
\Blue
\multiput(38,22)(4,0){2}{\line(0,1){4}}
\multiput(38,22)(0,4){2}{\line(1,0){4}}
\multiput(30,14)(4,0){4}{\line(0,1){8}}
\multiput(30,14)(0,4){3}{\line(1,0){12}}
\multiput(18,6)(4,0){7}{\line(0,1){8}}
\multiput(18,6)(0,4){3}{\line(1,0){24}}
\multiput(14,2)(4,0){8}{\line(0,1){4}}
\multiput(14,2)(0,4){2}{\line(1,0){28}}
\Black
\put(-3,33.5){$i_1$}\put(-3,29.5){$i_2$}\put(-3,25.5){$i_3$}\put(-3,21.5){$i_4$}\put(-3,17.5){$i_5$}
\put(-3,13.5){$i_6$}\put(-3,9.5){$i_7$}\put(-3,5.5){$i_8$}\put(-3,1.5){$i_9$}
\put(1,37.5){$r_1$}\put(5,37.5){$r_2$}\put(9,37.5){$r_3$}\put(13,37.5){$r_4$}\put(17,37.5){$r_5$}
\put(21,37.5){$r_6$}\put(25,37.5){$r_7$}\put(29,37.5){$r_8$}\put(33,37.5){$r_9$}\put(37,37.5){$r_{10}$}\put(41,37.5){$r_{11}$}
	\end{picture}}
	\vskip6pt
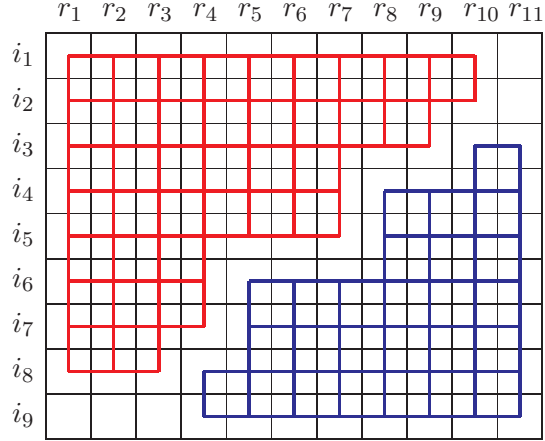
\centerline{Figure 2. \sl A Step Diagram}
\vskip6pt

As an example, for $(m,n)=(4,4)$, we have 6 possible diagrams corresponding to the six terms in the known representations of the $(4,4)$ split helicity amplitude \cite{RMV1, BCF}; these are:

\centerline{
\begin{picture}(20,20)
\linethickness{0.075mm}
\multiput(0,0)(4,0){5}{\line(0,1){16}}
\multiput(0,0)(0,4){5}{\line(1,0){16}}
\Red
\linethickness{0.3mm}
\multiput(2,6)(4,0){3}{\line(0,1){8}}
\multiput(2,6)(0,4){3}{\line(1,0){8}}
\Black
\end{picture}
\begin{picture}(20,20)
\linethickness{0.075mm}
\multiput(0,0)(4,0){5}{\line(0,1){16}}
\multiput(0,0)(0,4){5}{\line(1,0){16}}
\Red
\linethickness{0.3mm}
\multiput(2,10)(4,0){3}{\line(0,1){4}}
\multiput(2,10)(0,4){2}{\line(1,0){8}}
\multiput(2,6)(4,0){2}{\line(0,1){4}}
\multiput(2,6)(0,4){2}{\line(1,0){4}}
\Blue
\multiput(10,2)(4,0){2}{\line(0,1){4}}
\multiput(10,2)(0,4){2}{\line(1,0){4}}
\Black
\end{picture}
\begin{picture}(20,20)
\linethickness{0.075mm}
\multiput(0,0)(4,0){5}{\line(0,1){16}}
\multiput(0,0)(0,4){5}{\line(1,0){16}}
\Red
\linethickness{0.3mm}
\multiput(2,6)(4,0){2}{\line(0,1){8}}
\multiput(2,6)(0,4){3}{\line(1,0){4}}
\Blue
\multiput(10,2)(4,0){2}{\line(0,1){8}}
\multiput(10,2)(0,4){3}{\line(1,0){4}}
\Black
\end{picture}
}

\centerline{
\begin{picture}(20,20)
\linethickness{0.075mm}
\multiput(0,0)(4,0){5}{\line(0,1){16}}
\multiput(0,0)(0,4){5}{\line(1,0){16}}
\Blue
\linethickness{0.3mm}
\multiput(6,2)(4,0){3}{\line(0,1){8}}
\multiput(6,2)(0,4){3}{\line(1,0){8}}
\Black
\end{picture}
\begin{picture}(20,20)
\linethickness{0.075mm}
\multiput(0,0)(4,0){5}{\line(0,1){16}}
\multiput(0,0)(0,4){5}{\line(1,0){16}}
\Red
\linethickness{0.3mm}
\multiput(2,10)(4,0){2}{\line(0,1){4}}
\multiput(2,10)(0,4){2}{\line(1,0){4}}
\Blue
\multiput(6,2)(4,0){2}{\line(0,1){4}}
\multiput(6,2)(0,4){2}{\line(1,0){4}}
\multiput(10,2)(4,0){2}{\line(0,1){8}}
\multiput(10,2)(0,4){3}{\line(1,0){4}}
\Black
\end{picture}
\begin{picture}(20,20)
\linethickness{0.075mm}
\multiput(0,0)(4,0){5}{\line(0,1){16}}
\multiput(0,0)(0,4){5}{\line(1,0){16}}
\Red
\linethickness{0.3mm}
\multiput(2,10)(4,0){3}{\line(0,1){4}}
\multiput(2,10)(0,4){2}{\line(1,0){8}}
\Blue
\multiput(6,2)(4,0){3}{\line(0,1){4}}
\multiput(6,2)(0,4){2}{\line(1,0){8}}
\Black
\end{picture}
}
	\vskip6pt
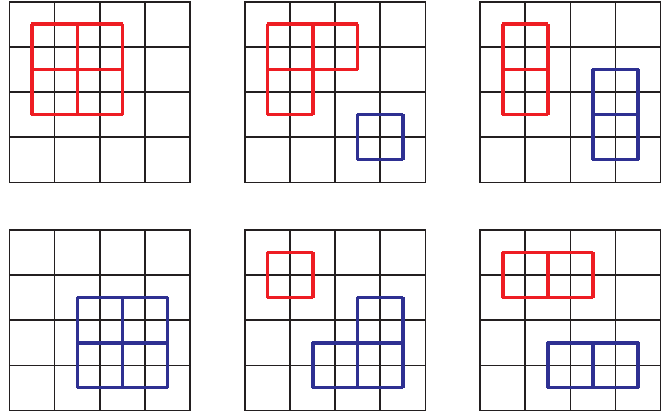
\centerline{Figure 3. \sl Step Diagrams for the $(4,4)$ Split Helicity Amplitude.}
\vskip6pt

To turn the relation (\ref{bigMM}) into a recurrence relation that expresses $\M_{mn}$ in terms of $\M_{m-1,n}$ and $\M_{m,n-1}$ we need to show how the appropriate set of conditions $f_{ab}=0$ enable the reduction of the $m+n$ equations (\ref{L1}) and (\ref{L2}) for the $mn$ variables $c_{lu}$, $l\in\P, u\in\N$,  to a similar $m+n-1$ equations for a reduced set of variables, and, in so doing, to specify the $m+n-1$ momenta that are the arguments of reduced amplitudes $\M_{m-1,n}$ and $\M_{m,n-1}$. We shall explain how to do this in the next section. Then we need to evaluate the Jacobians involved in the calculation of the residues in  (\ref{bigMM}) and in relating the integration variables that remain after taking the residue to those appropriate to $\M_{m-1,n}$ and $\M_{m,n-1}$. This we do in section \ref{Jacobians}.

\section{\bf  Reducing the Equations for $c_{lu}$}
\label{GenSol}

For a specific step diagram, the conditions (\ref{VW}), together with (\ref{L1}) and (\ref{L2}), enable us to evaluate the link variables $c_{lu}$ iteratively.

(a) Suppose that the step diagram has $m_1-1$ rows of maximal length, {\it i.e.} $n-2$ boxes, and that $m_1>1$. Let $\Sc=\{i_2, \ldots, i_{m_1}\}$ and write $i=i_1, j= i_m, r=r_1, s=r_n$ as usual. Then, for this diagram,
\be c^{ik}_{tu}=0,\qquad k\in\Sc, \quad t, u\in\N,\; t,u\ne s,\label{cS}\ee
and, from (\ref{L1}),
\be
c_{iu}\pi_k-c_{ku}\pi_i=\sum_{t\in\N}c^{ik}_{ut}\pi_t
=c^{ik}_{us}\pi_s,\qquad k\in\Sc,\quad u\in\N, u\ne s.
\ee
Using these equations, we can express $c_{ku}$ in terms of $c_{iu}$, $k\in\Sc, u\in\N$:

\begin{align}
{c_{ku}\over \langle k,s\rangle}&={c_{iu}\over \langle i,s\rangle},
\qquad k\in\Sc,\quad u\in\N, u\ne s,\label{cku}\\
{c_{ks}\over \langle k,s\rangle}&={c_{is}\over \langle i,s\rangle}
-{\langle k,i\rangle\over\langle k,s\rangle \langle i,s\rangle},
\qquad k\in\Sc.\label{cks}
\end{align}
Then, for $t\in\N, t\ne s$, we can write (\ref{L2})
\begin{align}
- \bpi_t
&=\left[\sum_{k\in\bS}\bpi_k{\langle k,s\rangle\over \langle i,s\rangle}
\right]c_{it}+\sum_{k\in\P\sim\bS}\bpi_kc_{kt},\cr
&=\bpi_i'c_{it}+\sum_{k\in\P\sim\bS}\bpi_kc_{kt},
\label{n1}
\end{align}
where $\bS=\Sc\cup\{i,s\}$ and  the shifted $\bpi_i'$ is defined as
\be\bpi_i'=\sum_{k\in\bS}\bpi_k{\langle k,s\rangle\over \langle i,s\rangle}
\equiv {1\over\langle i,s \rangle}\P_{\bS}\pi_s;\label{bpii}\ee 
further
\begin{align}
- \bpi_s&=\sum_{k\in\P}\bpi_kc_{ks}
=\left[\sum_{k\in\bS}\bpi_k{\langle k,s\rangle\over \langle i,s\rangle}\right]
c_{is}+\left[\sum_{k\in\Sc}\bpi_k{\langle k,i\rangle\over \langle s,i\rangle}
\right]+\sum_{k\in\P\sim\bS}\bpi_kc_{ks},\cr
- \bpi_s'&=\bpi_i'c_{is}+\sum_{k\in\P\sim\bS}\bpi_kc_{ks},\label{n2}\cr
\end{align}
with
\be\bpi_s'=\sum_{k\in\bS}\bpi_k{\langle k,i\rangle\over \langle s,i\rangle}
\equiv {1\over\langle s,i \rangle}\P_{\bS}\pi_i.\label{bpis}\ee
So taking the $m'=m-|\Sc|=m-m_1+1$ equations (\ref{L1}) for $k\in\P\sim\Sc$ 
and the new $n$ equations (\ref{n1}) and (\ref{n2}), we have a smaller 
system of equations with $(m,n)$ replaced by $(m',n)$, that is we have 
reduced the number of equations by $m_1-1$.

Further, we note that
\be\pi_i(\bpi_i')^T+\pi_s(\bpi_s')^T=\sum_{k\in\bS}\left[\pi_s{\langle k,i
\rangle\over \langle s,i\rangle}+\pi_i{\langle k,s\rangle\over 
\langle i,s\rangle}\right](\bpi_k)^T=\sum_{k\in\bS}\pi_k(\bpi_k)^T
\equiv\P_{\bS},\label{PS}\ee
as can be verified by taking the angle bracket with $\pi_i$ and $\pi_s$. Squaring this equation,
\be\langle i,s\rangle [i',s']=\left[p_i+\sum_{k\in\Sc}p_k+p_s\right]^2
=p_{\bS}^2.\label{PS2}\ee

(b) If the step diagram has  no rows of maximal length ({\it i.e.} $m_1=1$), it must have some number of columns of
maximal height, say $n-n_1$, $n>n_1$. Let $\T=\{r_{n_1},\ldots,r_{n-1}\}$
Then, for this diagram,
\be c^{kl}_{ts}=0,\qquad t\in\T, \quad k,l\in\P,\; k,l\ne i,\label{cT}\ee
and, from (\ref{L2}),
\be
\bpi_t  c_{ks}-\bpi_s c_{kt}=\sum_{l\in\P}\bpi_l c^{kl}_{ts}=\bpi_i 
c^{ki}_{ts},\qquad t\in\T,\quad k\in\P, k\ne i.
\ee
From these relations, as in (a), we can replace (\ref{L1}), (\ref{L2}) with a reduced set of equations

\begin{align}
\pi_k&=\sum_{t\in\N\sim\bT}c_{kt}\pi_t+c_{ks}\pi_s',\qquad k\in\P,k\ne i,\label{p1}\\
\pi_i'&=\sum_{t\in\N\sim\bT}c_{it}\pi_t+c_{is}\pi_s', \label{p2}
\end{align}
where $\bT=\T\cup\{i,s\} $ and the shifted $\pi'_s$ and $\pi'_i$ are defined by 
\begin{align}\pi_s'&=\sum_{t\in\bT}\pi_t{[t,i]\over[s,i]}\equiv {1\over[ s,i ]}
\P_{\bT}\bpi_i, \label{pis}\\
\pi_i'&=\sum_{t\in\bT}\pi_t{[t,s]\over[i,s]}\equiv {1\over[ i,s ]}\P_{\bT}
\bpi_s\label{pii}.
\end{align}
So taking the $n'=n-|\T|=n_1$ equations (\ref{L2}) for $t\in\N\sim\T$ and the new $m$ 
equations (\ref{p1}) and (\ref{p2}), we have a smaller system of equations 
with $(m,n)$ replaced by $(m,n_1)$, that is we have reduced the number of 
equations by $n-n_1$.

As before, in (\ref{PS}) and (\ref{PS2}), 
\be\pi_i'\bpi_i^T+\pi_s'\bpi_s^T=\sum_{t\in\bT}\pi_t(\bpi_t)^T\equiv
\P_{\bT},\label{PT}\ee
and 
\be\langle i',s'\rangle [i,s]=\left[p_i+\sum_{t\in\T}p_t+p_s\right]^2
= p_{\bT}^2.\label{PT2}\ee

\section{\bf The Recurrence Relation}

In (\ref{bigMM}), $\M_{mn}$ is written as the sum of two contributions and, from (\ref{recc1}), the first of which is

\be\hat K_{mn}\oint 
{\hat F_{m-1,n}(c) \over c_{11} f_{1,n-1}}\left|{\partial \, f\over\partial\hat\beta}\right|^{-1}\prod_{b=1}^{n-1} c_{2b}
\prod_{a=3}^{m-1}\prod_{b=2}^{n-1}{d\hat\beta_{ab}\over C_{ab}}. \ee
Using (\ref{simform}), this becomes
\be
{\langle r_n,i_2\rangle^{n-1}   [i_2,i_3]^{n-2}  \over\langle r_n|P_{i_1i_2}|i_3]^{n-2}
\langle i_1,i_2\rangle\langle r_n,i_1\rangle} \hat K_{m-1,n}\oint \hat F_{m-1,n}(c) 
\prod_{a=3}^{m-1}\prod_{b=2}^{n-1}{d\hat\beta_{ab}\over C_{ab}} 
\ee

If we use the conditions $ f_{1b}=0$, $1\leq b\leq n-2$, 
to eliminate $\pi_{i_1}, \bpi_{i_1},  c_{1b}, 1\leq b\leq n$, 
to define a new system as in section \ref{GenSol}(a), consisting of
\be\pi_{i_2},\;\bpi_{i_2}'={1\over \langle i_2,r_n\rangle}\P_{\bS}\pi_{r_n}; \quad
\pi_{i_a},\;\bpi_{i_a}, \,3\leq a\leq m;\ee
\be  \pi_{r_b},\;\bpi_{r_b}, \,
1\leq b\leq n-1;\;\quad
\pi_{r_n},\,\bpi_{r_n}'={1\over \langle r_n,i_2\rangle}\P_{\bS}\pi_{i_2}; \ee
where $\bS=\{i_1,i_2,r_n\}$, and $ c_{ab}$, $2\leq a\leq m, 1\leq b\leq n$, 
and then use this system to define a set of contiguous $\beta$'s, 
$\hat\beta'_{ab}$, 
$3\leq a\leq m-1, 2\leq b\leq n-1$,  these will not be the same as the 
corresponding $\hat\beta_{ab}$. So we have to incorporate the corresponding Jacobian (\ref{jac3})

\begin{align}
{\langle r_n,i_2\rangle^{n-1}   [i_2,i_3]^{n-2}  \over\langle r_n|P_{i_1i_2}|i_3]^{n-2}
\langle i_1,i_2\rangle\langle r_n,i_1\rangle}&\hat K_{m-1,n} \oint \hat F_{m-1,n}(c) 
\left|{\partial \hat\beta'\over\partial\hat\beta}\right|^{-1}\prod_{a=3}^{m-1}\prod_{b=2}^{n-1}{d\hat\beta_{ab}'\over C_{ab}} 
\cr
&\hskip-10truemm={\langle r_n,i_2\rangle  \over\langle r_n,i_1\rangle\langle i_1,i_2\rangle}
\hat K_{m-1,n}\oint \hat 
F_{m-1,n}(c) \prod_{a=3}^{m-1}\prod_{b=2}^{n-1}{d\hat\beta_{ab}'
\over C_{ab}}\cr
&\hskip-10truemm={\langle r_n,i_2\rangle  \over\langle r_n,i_1\rangle\langle i_1,i_2\rangle}
\M_{m-1,n}.\end{align}

Making a similar evaluation of the second contribution $\M_{mn}$ in (\ref{bigMM}), we have
\be\M_{mn}={\langle r_n,i_2\rangle\over\langle r_n,i_1\rangle\langle i_1,i_2
\rangle}\M_{m-1,n}+{[r_{n-1},i_1]\over[r_{n-1},r_n][r_{n},i_1]}\M_{m,n-1}.
\label{recur2}\ee

Note this is the BCFW recursion relation restricted to 
split helicity amplitudes \cite{BCF,BCFW}, where
the shifted momenta $\bpi_i', \bpi_s'$ in $\M_{m-1,n}$ are given by
(\ref{bpii}) and (\ref{bpis}), and the shifted momenta $\pi_i', \pi_s'$ in
 $\M_{m,n-1}$ are given by
(\ref{pii}) and (\ref{pis}), correspond to the BCFW reference
momenta. So we have a complete proof that the split amplitudes
for the twistor string agree with gauge theory. 

Note, from (\ref{mhv}),
\begin{alignat}{3}
 \M_{m2} & = {K_{m2}\over c_{11} c_{m2}} \prod_{a=1}^{m-1}{1\over f_{a1}}
& & ={\langle r_1,r_2\rangle^3\over \langle r_2,i_1\rangle\langle i_m,r_1\rangle}\prod_{a=1}^{m-1}{1\over\langle i_a,i_{a+1}\rangle}\label{mhv2}\\
\M_{2n} &= {K_{2n}\over c_{11} c_{2n}} \prod_{b=1}^{n-1}{1\over f_{1b}}
&  &= {[i_1,i_2]^3\over [r_n,i_1][i_2,r_1]}\prod_{b=1}^{n-1}{1\over[r_b,r_{b+1}]},\label{mhv3}
\end{alignat}
using
\begin{alignat}{3}
c_{a1}&={\langle i_a,r_2\rangle\over\langle r_1,r_2\rangle},&\quad
c_{a2}&=-{\langle i_a,r_1\rangle\over\langle r_1,r_2\rangle},&\quad
f_{a1}&={\langle i_a,i_{a+1}\rangle\over\langle r_1,r_2\rangle},\cr
\noalign{\hskip10.2truemm from (\ref{L1}) in (\ref{mhv2}), and }
c_{1b}&=-{[r_b,i_2]\over [i_1,i_2]},&\qquad 
c_{2b}&={[r_b,i_1]\over [i_1,i_2]},&\qquad 
f_{1b}&={[r_b,r_{b+1}]\over [i_1,i_2]},\cr
\noalign{\hskip10.2truemm from (\ref{L2}) in (\ref{mhv3}). }
\nonumber\end{alignat}

\vskip-22pt

We consider applying the recurrence relation $\ell$ times and we calculate the term that comes from removing the top row each time. This involves 
$\M_{m-\ell,n}$ with momenta 
\be \pi_{i_{\ell+1}},\ldots,\pi_{i_m}; \;\pi_{r_1},\ldots, \pi_{r_n}; \;
 \bpi_{i_{\ell+1}}', \bpi_{i_{\ell+2}},\ldots,\bpi_{i_m};\; \bpi_{r_1},\ldots, \bpi_{r_n}^{(\ell+1)},\ee
 where $\bpi_{i_1}'\equiv\bpi_{i_1}$ and $\bpi_{r_n}^{(1)}=\bpi_{r_n}$, and the iterative relation
 \be 
  \bpi_{i_{\ell+1}}'=\bpi_{i_\ell}'{\langle i_\ell, r_n\rangle\over \langle i_{\ell+1},r_n\rangle}
  +\bpi_{i_{\ell+1}},\quad
\bpi_{r_n}^{(\ell+1)}=\bpi_{i_\ell}'{\langle i_\ell,i_{\ell+1}\rangle\over \langle i_{\ell+1},r_n\rangle}
 -\bpi_{r_n}^{(\ell)}
\ee
has the solution
\be\bpi_{i_\ell}'={1\over \langle i_\ell,r_n\rangle}\P_{\bS_\ell}\pi_{r_n},
\quad\bpi_{r_n}^{(\ell)}={1\over \langle r_n,i_\ell\rangle}\P_{\bS_\ell}
\pi_{i_\ell}, \ee
where $\bS_\ell=\{r_n,i_a: 1\leq a\leq\ell\}$.
Thus we may build up $\bS_\ell$ adding one particle at a time.    
and it follows that the contribution to $\M_{mn}$ with $ f_{ab}=0, 1\leq a 
\leq\ell, 1\leq b\leq n-2,$ is
\be
{\langle r_n,i_{\ell +1}\rangle   \over\langle r_n,i_1\rangle}\prod_{a=1}^\ell
{1\over\langle i_a,i_{a+1}\rangle}\M_{m-\ell,n}.\label{onebyone}\ee
Similarly the contribution to $\M_{mn}$ with $ f_{ab}=0, 2\leq a 
\leq m-1, n-\ell\leq b\leq n-1,$ is
\be
{[ r_{n-\ell},i_1]   \over[ r_n,i_1]}\prod_{b=n-\ell}^{n-1}
{1\over[r_b,r_{b+1}]}\M_{m,n-\ell},\label{onebyone2}\ee
where
$\M_{m,n-\ell}$ with momenta 
\be \pi_{i_{1}}^{(\ell+1)},\ldots,\pi_{i_m}; \;\pi_{r_1},\ldots, \pi_{r_{n-\ell}}'; \;
 \bpi_{i_{1}}, \bpi_{i_{\ell+2}},\ldots,\bpi_{i_m};\; \bpi_{r_1},\ldots, \bpi_{r_{n-\ell}},\ee
 \be\pi_{i_1}^{(\ell+1)}={1\over [ i_1,r_{n-\ell}]}\P_{\bT_{\ell+1}}\bpi_{r_{n-\ell}},
\quad\pi_{r_{n-\ell}}'={1\over [ r_{n-\ell},i_1]}\P_{\bT_{\ell+1}}
\bpi_{i_1},\ee
and $\bT_\ell=\{r_b,i_1: n-\ell< b\leq n\}$.

In the next section we extend these results to obtain the expression for the contribution associated with a general step diagram.

\section{\bf General Formula.}\label{sect:GenForm}
\label{GenForm}

In this section we evaluate the contribution to the $(m,n)$ split helicity amplitude $\M_{mn}$ associated with a general step diagram using the analysis of the previous sections. We can describe such a diagram as follows: suppose it has $m_1-1$ rows of maximal length, {\it i.e.}  $n-2$, (so that $m_1=1$ if there are no such rows); $n-n_1$ columns of the maximal permitted height given $m_1$, {\it i.e.} $m-m_1-1$; $m_2-m_1$ rows of the next maximal length given $n_1$, {\it i.e.} $n_1-2$; $n_1-n_2$ columns of height $m-m_2-1$;    and so on until we reach $n_{p-1}-n_p$ columns of height $m-m_p-1$  and, finally, $m-m_p-1$ rows of length $n_p-2$.

Associated with this step diagram we have the conditions:
\be f_{ab}=0,\quad m_{q-1}\leq a<m_q,\quad 1\leq b <  n_{q-1}-1, \quad 1\leq q\leq p+1,\label{mfab}\ee
\be f_{ab}=0,\quad m_q< a<m,\quad  n_q\leq b <n_{q-1}, \quad 1\leq q\leq p,\label{nfab}\ee
where $m_0=1, n_0=n, m_{p+1}=m-1$.

We evaluate the contribution to $\M_{mn}$ by successively reducing the amplitude using (\ref{onebyone}) and  (\ref{onebyone2}).
This process is illustrated in Figure 4 for the step diagram specified by 
$$m=6, \;n=8,\;p=2, \;m_1=2, \;m_2=3,\;m_3=5, \;n_1=6, \;n_2=3.$$

\vskip10pt 

\setlength{\unitlength}{0.8mm}

\centerline{\begin{picture}(33,25)
\linethickness{0.075mm}
\multiput(0,0)(4,0){9}{\line(0,1){24}}
\multiput(0,0)(0,4){7}{\line(1,0){32}}
\Red
\linethickness{0.3mm}
\multiput(2,18)(4,0){7}{\line(0,1){4}}
\multiput(2,18)(0,4){2}{\line(1,0){24}}
\multiput(2,14)(4,0){5}{\line(0,1){4}}
\multiput(2,14)(0,4){2}{\line(1,0){16}}
\multiput(2,6)(4,0){2}{\line(0,1){8}}
\multiput(2,6)(0,4){3}{\line(1,0){4}}
\Blue
\multiput(22,2)(4,0){3}{\line(0,1){12}}
\multiput(22,2)(0,4){4}{\line(1,0){8}}
\multiput(10,2)(4,0){4}{\line(0,1){8}}
\multiput(10,2)(0,4){3}{\line(1,0){12}}
\Black
\end{picture} 
\begin{picture}(6,25)
\put(0,12){$\rightarrow$}
\end{picture}
\begin{picture}(33,25)
\linethickness{0.075mm}
\multiput(0,4)(4,0){9}{\line(0,1){20}}
\multiput(0,4)(0,4){6}{\line(1,0){32}}
\Red
\linethickness{0.3mm}
\multiput(2,18)(4,0){5}{\line(0,1){4}}
\multiput(2,18)(0,4){2}{\line(1,0){16}}
\multiput(2,10)(4,0){2}{\line(0,1){8}}
\multiput(2,10)(0,4){3}{\line(1,0){4}}
\Blue
\multiput(22,6)(4,0){3}{\line(0,1){12}}
\multiput(22,6)(0,4){4}{\line(1,0){8}}
\multiput(10,6)(4,0){4}{\line(0,1){8}}
\multiput(10,6)(0,4){3}{\line(1,0){12}}
\Black
\end{picture} 
\begin{picture}(6,25)
\put(0,12){$\rightarrow$}
\end{picture}
\begin{picture}(25,25)
\linethickness{0.075mm}
\multiput(0,4)(4,0){7}{\line(0,1){20}}
\multiput(0,4)(0,4){6}{\line(1,0){24}}
\Red
\linethickness{0.3mm}
\multiput(2,18)(4,0){5}{\line(0,1){4}}
\multiput(2,18)(0,4){2}{\line(1,0){16}}
\multiput(2,10)(4,0){2}{\line(0,1){8}}
\multiput(2,10)(0,4){3}{\line(1,0){4}}
\Blue
\multiput(10,6)(4,0){4}{\line(0,1){8}}
\multiput(10,6)(0,4){3}{\line(1,0){12}}
\Black
\end{picture} 
\begin{picture}(6,25)
\put(0,14){$\rightarrow$}
\end{picture}
\begin{picture}(25,25)
\linethickness{0.075mm}
\multiput(0,8)(4,0){7}{\line(0,1){16}}
\multiput(0,8)(0,4){5}{\line(1,0){24}}
\Red
\linethickness{0.3mm}
\multiput(2,14)(4,0){2}{\line(0,1){8}}
\multiput(2,14)(0,4){3}{\line(1,0){4}}
\Blue
\multiput(10,10)(4,0){4}{\line(0,1){8}}
\multiput(10,10)(0,4){3}{\line(1,0){12}}
\Black
\end{picture} 
\begin{picture}(6,25)
\put(0,14){$\rightarrow$}
\end{picture}
\begin{picture}(13,25)
\linethickness{0.075mm}
\multiput(0,8)(4,0){4}{\line(0,1){16}}
\multiput(0,8)(0,4){5}{\line(1,0){12}}
\Red
\linethickness{0.3mm}
\multiput(2,14)(4,0){2}{\line(0,1){8}}
\multiput(2,14)(0,4){3}{\line(1,0){4}}
\Black
\end{picture}
\begin{picture}(6,25)
\put(0,18){$\rightarrow$}
\end{picture}
\begin{picture}(10,25)
\linethickness{0.075mm}
\multiput(0,16)(4,0){4}{\line(0,1){8}}
\multiput(0,16)(0,4){3}{\line(1,0){12}}
\Black
\end{picture} }
	\vskip6pt
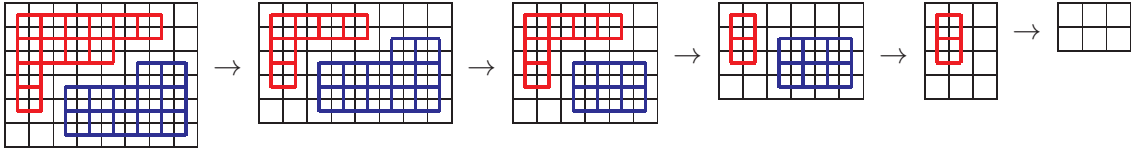
\centerline{Figure 4. \sl Reduction of a Step Diagram}
\vskip6pt

From (\ref{onebyone}), the contribution
to $\M_{m,n}$ for 
$$ f_{ab}=0,\quad 1\leq a<m_1,\, 1\leq b< n-1,$$
  is
$${\langle r_n, i_{m_1}\rangle\over \langle r_n,i_1\rangle \langle i_1,i_2\rangle}
\prod_{a=2}^{m_1-1} {1\over \langle i_a, i_{a+1}\rangle} \;\M_{m-m_1+1, n}
(i'_{m_1}, i_{m_1+1}, \ldots i_m, r_1,\ldots, r_{n-1},r'_n),$$
with
$$[i'_{m_1}]={1\over \langle i_{m_1},r_n\rangle}\P_{\bS_1}| r_n\rangle, \quad 
|i'_{m_1}\rangle=|i_{m_1}\rangle,\quad
|r'_n]={1\over \langle r_n, i_{m_1}\rangle}\P_{\bS_1}| i_{m_1}\rangle, \quad
|r'_n\rangle =|r_n\rangle,$$
$\bS_1\equiv\bS_{m_1}=\{r_n, i_1, \ldots, i_{m_1-1}, i_{m_1}\},$
 thus eliminating $i_1, \ldots, i_{m_1-1}$.

From (\ref{onebyone2}), the contribution to $\M_{m-m_1+1, n}
(i'_{m_1}, i_{m_1+1}, \ldots i_m, r_1,\ldots, r_{n-1},r'_n)$
for 
$$ f_{ab}=0,\quad m_1< a<m,\,  n_1\leq b < n,$$
 is
$${[i'_{m_1}, r_{n_1}]\over [i'_{m_1},r'_n] [r'_n ,r_{n-1}]}
\prod_{b={n_1+1}}^{n-1} {1\over [r_b, r_{b-1}]} \;\M_{m-m_1+1, n_1}
(i''_{m_1}, i_{m_1+1}, \ldots, i_m, r_1,\ldots, r_{n_1-1}, r'_{n_1}),$$
with
$$|i''_{m_1}\rangle={1\over [i'_{m_1}, r_{n_1}]}\P_{\bT_1}| r_{n_1}
\rangle,\quad |i''_{m_1}]=|i'_{m_1}],\quad |r'_{n_1}\rangle ={1\over  [r_{n_1},i'_{m_1} ]}\P_{\bT_1}| i'_{m_1}],\quad
|r'_{n_1} ] =|r_{n_1}],$$
$\bT_1\equiv\bT_{n_1}=\{r_{n_1}, r_{n_1+1},\ldots, r'_n, i'_{m_1}\}$, thus eliminating $r_{n_1+1}, \ldots, r'_n$.

Proceeding inductively, from (\ref{onebyone}), the contribution
to
$$\M_{m-m_{q}+1, n_q}
(i''_{m_{q}}, i_{m_{q}+1}, \ldots, i_m, r_{q},\ldots, r_{n_{q}-1}, r'_{n_{q}})$$
for 
$$ f_{ab}=0,\quad m_q\leq a<m_{q+1},\, 1\leq b< n_q-1, $$
is
$${ \langle r'_{n_q}, i_{m_{q+1}}\rangle \over \langle r'_{n_q} ,i''_{m_q}\rangle
\langle i''_{m_q} ,i_{m_q+1}\rangle}
\prod_{a=m_q+1}^{m_{q+1}-1} {1\over \langle i_a, i_{a+1}\rangle} \;\M_{m-m_{q+1}+1, n_q}
(i'_{m_{q+1}}, i_{m_{q+1}+1}, \ldots, i_m, r_q,\ldots, r_{n_q-1}, r''_{n_q}),$$
with
$$ [i'_{m_{q+1}}]={1\over \langle i_{m_{q+1}},r'_{n_q}\rangle}
\P_{\bS_{q+1}}| r'_{n_q}\rangle,\quad |i'_{m_{q+1}}\rangle=|i_{m_{q+1}}\rangle,$$
$$ |r''_{n_q}]=
{1\over\langle r'_{n_q}, i_{m_{q+1}}\rangle}\P_{\bS_{q+1}} | i_{m_{q+1}}\rangle,
\quad |r''_{n_q}\rangle =|r'_{n_q}\rangle,$$
$\bS_{q+1}\equiv\bS_{m_{q+1}}=\{r'_{n_q}, i''_{m_q}, \ldots, i_{m_{q+1}-1}, i_{m_{q+1}}\},$  thus eliminating
$i''_{m_q}, \ldots, i_{m_{q+1}-1}.\hfil$

Proceeding inductively, from (\ref{onebyone2}), the contribution to 
$$\M_{m-m_{q+1}+1, n_q}(i'_{m_{q+1}}, i_{m_{q+1}+1}, \ldots, i_m, r_1,\ldots, r_{n_q-1}, r''_{n_q})$$ 
for 
$$ f_{ab}=0,\quad m_{q+1}< a<m,\quad  n_{q+1}\leq b < n_q,$$
is
$${[i'_{m_{q+1}} r_{n_{q+1}}]\over [i'_{m_{q+1}}r''_{n_q}] [r''_{n_q} r_{n_q-1}]}
\prod_{b={n_{q+1}+1}}^{n_q-1} {1\over [r_b r_{b-1}]} \;\M_{m-m_q+1, n_{q+1}}
(i''_{m_{q+1}}, i_{m_{q+1}+1}, \ldots, i_m, r_1,\ldots, r_{n_{q+1}-1}, r'_{n_{q+1}}),$$
with
$$|i''_{m_{q+1}}\rangle={1\over [i'_{m_{q+1}}, r_{n_{q+1}}]}\P_{\bT_{q+1}}| r_{n_{q+1}}
\rangle,\quad |i''_{m_{q+1}}]=|i'_{m_{q+1}}]$$
$$|r'_{n_{q+1}}\rangle ={1\over  [r_{n_{q+1}},i'_{m_{q+1}} ]}\P_{\bT_{q+1}}
| i'_{m_{q+1}}],\quad
|r'_{n_{q+1}} ] =|r_{n_{q+1}}],$$
$\bT_{q+1}\equiv\bT_{n_{q+1}}=\{r_{n_{q+1}}, r_{n_{q+1}+1}, \ldots, r''_{n_q}, i'_{m_{q+1}}\},$ thus eliminating $r_{n_{q+1}+1}, \ldots, r''_{n_q}$.

It follows that
\begin{align}
|i_{m_q}''\rangle&={1\over [ i_{m_q}',r_{n_q}]}\P_{\bT_q}|\ r_{n_q}],
\qquad|i_{m_{q}}'\rangle=|i_{m_{q}}\rangle,\cr
|i_{m_q}'']=|i_{m_q}']
&={1\over \langle i_{m_{q}},r_{n_{q-1}}'\rangle[r_{n_{q-1}, i_{m_{q-1}}'}]
\ldots  \langle i_{m_1},r_n\rangle}\P_{\bS_{q}}\P_{\bT_{q-1}}\ldots
\P_{\bS_1}|\ r_n\rangle,\label{rel1}\\
|r_{n_{q}}'']
&={1\over \langle r_{n_{q}}', i_{m_{q+1}}\rangle}\P_{\bS_1}|i_{m_{q+1}}
\rangle,\qquad |r_{n_q}']=|r_{n_q}],\cr 
|r_{n_{q}}''\rangle=|r_{n_{q}}'\rangle
&={1\over [r_{n_q}, i_{m_q}'] \langle i_{m_{q}},r_{n_{q-1}}'\rangle\ldots
\langle i_{m_{1}},r_{n}\rangle}\P_{\bT_q}\P_{\bS_{q}}\ldots\P_{\bS_{1}}
|r_{n}\rangle\label{rel2}.\end{align}

So the factor associated with $\bS_q$ is
$${[ i_{m_{q-1}}',r_{n_{q-1}}]^2\langle 
r_{n_{q-1}}',i_{m_q}\rangle\over p_{\bT_{q-1}}^2\langle i_{m_{q-1}+1} | 
\P_{\bT_{q-1}}|\ r_{n_{q-1}}]}\prod_{a=m_{q-1}+1}^{m_q-1}{1\over \langle 
i_a,i_{a+1} \rangle}$$
and the factor associated with $\bT_q$ is
$$ {\langle r_{n_{q-1}}',
i_{m_q}\rangle^2[i_{m_q}',r_{n_{q}}] \over p_{\bS_q}^2[r_{n_{q-1}-1}|
\P_{\bS_q}|i_{m_{q}}\rangle }  
\prod_{b=n_{q}+1}^{n_{q-1}-1}{1\over [r_{b},r_{b-1}]}.$$

Combining these factors for $1\leq q\leq p$, we obtain
\begin{align}
&\prod_{q=1}^p\left[{[ i_{m_{q-1}}',r_{n_{q-1}}]^2\langle 
r_{n_{q-1}}',i_{m_q}\rangle\langle r_{n_{q-1}}',i_{m_q}\rangle^2[i_{m_q}',r_{n_{q}}] \over p_{\bT_{q-1}}^2\langle i_{m_{q-1}+1} | 
\P_{\bT_{q-1}}|\ r_{n_{q-1}}]p_{\bS_q}^2[r_{n_{q-1}-1}|
\P_{\bS_q}|i_{m_{q}}\rangle } 
\prod_{a=m_{q-1}+1}^{m_q-1}{1\over \langle i_a,i_{a+1} \rangle}
\prod_{b=n_{q}+1}^{n_{q-1}-1}{1\over [r_{b},r_{b-1}]}\right] \cr
&\hskip50mm\times
\M_{m-m_p+1, n_p} (i'_{m_p}, i_{m_p+1}, \ldots, i_m r_1, \ldots, r_{n_p-1},
r'_{n_p}),\label{q1q2}\end{align}
We find the contribution to  $\M_{m-m_p+1, n_p} $ in (\ref{q1q2})
for
$$ f_{ab}=0,\quad m_p\le a< m_{p+1}\equiv m-1,\quad1\le b < n_p-1,$$
is  
\begin{align}
&{\langle r'_{n_p} i_{m-1}\rangle \over \langle r'_{n_p} i''_{m_p}\rangle
\langle i''_{m_p} i_{m_p +1}\rangle} \prod_{a=m_p+1}^{m-2}
{1\over \langle i_a i_{a+1}\rangle} \;\;\M_{2, n_p}(i'_{m-1}, i_m, r_1,
\ldots, r_{n_p-1}, r''_{n_p})\cr
&= {\langle r'_{n_p} i_{m-1}\rangle \over \langle r'_{n_p} i''_{m_p}\rangle
\langle i''_{m_p} i_{m_p +1}\rangle} \prod_{a=m_p+1}^{m-2}
{1\over \langle i_a i_{a+1}\rangle} \times
{[i_{m-1}',i_m]^3\over[i_{m-1}',r_{n_p}''][r_{n_p}'',r_{n_p-1}][r_1,i_m]} 
\prod_{b=2}^{n_{p}-1}{1\over [r_{b},r_{b-1}]}\cr
&= {[ i_{m_{p}}',r_{n_{p}}]^2\langle r_{n_{p}}',i_{m-1}\rangle\over  
p_{\bT_{p}}^2\langle i_{m_{p}+1} |\P_{\bT_{p}}|\ r_{n_{p}}]}
\prod_{a=m_{p}+1}^{m-2}{1\over \langle i_a,i_{a+1} \rangle}\times 
{\langle r_{n_{p}}',i_{m-1}\rangle^2[i_{m-1}',i_m]^3\over p_{\bS_{p+1}}^2
[r_{n_p-1}|\P_{\bS_{p+1}}|i_{m-1}\rangle [r_1,i_m]} \prod_{b=2}^{n_{p}-1}
{1\over [r_{b},r_{b-1}]},&\cr
\label{work}\end{align}
where
$$|i'_{m-1}] 
= {1\over \langle i_{m_{p+1}},r_{n_{p}}'\rangle [r_{n_{p}}, i_{m_{p}}']
\langle i_{m_{p}},r_{n_{p-1}}'\rangle [r_{n_{p-1}}, i_{m_{p-1}}']
\ldots \langle i_{m_1},r_n\rangle}
\P_{\bS_{p+1}}\P_{\bT_{p}}\ldots\P_{\bS_1}|\ r_n\rangle.$$
Let $\A=\{m_a: 1\leq a \leq p\}$, $\B=\{n_b: 1\leq b \leq p\}$.
Combining (\ref{q1q2}) with (\ref{work}), we have the general
formula for the contribution
to $\M_{m n}$ from the step diagram that we have considered:
\begin{align}
{1\over [r_1,i_m]\langle r_n,i_1 \rangle}  \prod_{1\leq a\leq m-2
\atop a\notin \A}&{1\over \langle i_a,i_{a+1} \rangle}\prod_{2\leq b\leq n-1
\atop b\notin \B}{1\over [r_{b},r_{b-1}]}\prod_{q=1}^{p}{1\over p^2_{\bT_q}
\langle i_{m_{q}+1} | \P_{\bT_{q}}|\ r_{n_{q}}]}\cr&
\times\prod_{q=1}^{p+1}
{1\over p^2_{\bS_q} [r_{n_{q-1}-1}|\P_{\bS_q}|i_{m_{q}}\rangle}
[i_m|\P_{\bS_{p+1}}\P_{\bT_{p}}\ldots\P_{\bS_1}|\ r_n\rangle^3
.\label{genform}\end{align}
This agrees with the gauge theory expression obtained in \cite{Britto}.

Note that $P_{\bS_q}=-P_{\bS_q'}$ and
$P_{\bT_q}=-P_{\bT_q'}$, where 
$$\bS_q'=\{i_{m_q+1},\ldots, i_m,r_1,
\ldots,r_{n_{q-1}-1}\},  \qquad\bT_q'=\{i_{m_q+1},\ldots, i_m,r_1,\ldots,
r_{n_{q}-1}\}.$$

[Although (\ref{genform}) directly addresses the case where the reduction process both 
begins and ends with an $\bS$, we can also obtain from the other cases in which the
reduction either begins or ends with a $\bT$, or both, by taking  $m_1 = 1$ and 
$\bS_1 = \{ r_n, i_1\}$ if the process begins with $\bT$ and  $n_p = 2$ and $\bS_{p+1} = \{i_m, r_1\}$
if it ends with $\bT$.]  

\section{\bf Examples}
We compute the $(4,4)$ and $(5,3)$ 
amplitudes from twistor string theory using 
(\ref{genform}). 
\vskip10pt

For $(4,4)$, the six possible
step diagrams are drawn in Figure 3.

a) \; The  lower left  diagram corresponds to
 $ f_{22} =  f_{23} =  f_{32} = f_{33} = 0$.
In the notation of section \ref{GenForm}, this diagram has $p=1$,
$n_0=n=4, m_1=1, n_1=2, m_2=3$ and
$\bS_1 = \{r_4,i_1\}$,\, $\bS_2 = \{i_4, r_1\}$,\,
$\bT_1 = \{r_2,r_3,r'_4,i'_1\}$, where
$\P_{\bT_1} = -\P_{\bT'_1}$ with
$\bT'_1 =  \{i_2, i_3, i_4, r_1\}.$
 
Then from (\ref{genform}) the contribution to $\M_{4,4}$ is
\begin{align}
&{1\over [r_1 i_4] \langle r_4i_1 \rangle}
{1\over \langle i_2 i_3\rangle [r_3 r_2]}
\,{[i_4| \P_{i_4r_1} \P_{i_2i_3i_4r_1} \P_{i_4r_1} | r_4\rangle^3
\over p^2_{i_2i_3i_4r_1} \langle i_2| \P_{i_2i_3i_4r_1} |r_2]
\,p^2 _{i_1r_4}\, [r_3|\P_{r_4 i_1}|i_1\rangle \;
p^2_{i_4r_1}\, [r_1|\P_{i_4 r_1} | i_3\rangle}\cr
&\qquad\qquad= -{\langle r_1| \P_{i_2i_3i_4} | i_1]^3\over
p^2_{i_2i_3i_4r_1} [r_2|\P_{i_3i_4r_1}|i_2\rangle [r_3r_4][r_4i_1][r_2r_3]
\langle i_2i_3\rangle \langle i_3i_4\rangle \langle i_4r_1\rangle}.
\nonumber\end{align}

b) The  lower right  diagram in  Figure 3  has
$ f_{11}= f_{12}= f_{32}= f_{33} = 0$.
This corresponds to $p=1$, $n_0=n=4, m_1=2, n_1=2, m_2=3$ and
 $\bS_1 = \{r_4,i_1,i_2\}$,\, $\bS_2 = \{i_4, r_1\}$,\,
$\bT_1 = \{r_2,r_3,r'_4,i'_2\}$, where
$\P_{\bT_1} = -\P_{\bT'_1}$ with
$\bT'_1 =  \{i_3, i_4, r_1\}$,
from formulae in section \ref{GenForm}.
The contribution to $\M_{4,4}$ yields
the second term listed in (\ref{finalfour}). 
\vskip 10pt

c) The  lower middle diagram in Figure 3 
has $ f_{11}= f_{23}= f_{32}= f_{33} = 0$,
which is described by
$p=2$, $n_0=n=4, m_1=1, n_1=3, m_2=2$ and
 $\bS_1 = \{r_4,i_1\}$,\, $\bS_2 = \{r'_3, i''_1, i_2\}$,\,
$\bS_3 = \{r_4, i_1\}$,\,
$\bT_1 = \{r_3,r'_4,i'_1\}$, $\bT_2 = \{r_2,r'_3,i'_2\}$,
where $\P_{\bS_2} = -\P_{\bS'_2}$ with
$\bS'_2 =  \{i_3, i_4, r_1, r_2\}$;
$\P_{\bT_1} = -\P_{\bT'_1}$ with
$\bT'_1 =  \{i_2, i_3, i_4, r_1, r_2\}$;
$\P_{\bT_2} = -\P_{\bT'_1}$ with
$\bT'_1 =  \{i_3, i_4, r_1\}$.
From (\ref{genform}), the contribution to $\M_{4,4}$ becomes the third term
in (\ref{finalfour}).

The  remaining  three diagrams are related to the
ones directly above them in the  Figure 3, 
by the symmetry flip $i_1\leftrightarrow r_4$,
$i_2\leftrightarrow r_3, i_3\leftrightarrow i_2, i_4\leftrightarrow r_1$.
Thus their contribution to $\M_{4,4}$ is found by the flip of the first
three terms, (while exchanging the angle with the square brackets). 

So we have
\begin{align}
\M_{4,4} = 
 &- {\langle r_1| \P_{i_2i_3i_4} | i_1]^3\over
p^2_{i_2i_3i_4r_1} [r_2|\P_{i_3i_4r_1}|i_2\rangle [r_3r_4][r_4i_1][r_2r_3]
\langle i_2i_3\rangle \langle i_3i_4\rangle \langle i_4r_1\rangle}\cr
& +  {\langle r_1 | \P_{i_3i_4} \P_{i_1i_2}| r_4\rangle^3\over
p^2_{i_3 i_4 r_1} p^2_{r_4 i_1 i_2} [r_2|\P_{i_4r_1}|i_3\rangle
[r_3 |\P_{r_4i_1} | i_2\rangle \,
[r_2r_3]\langle r_1 i_4\rangle \langle i_4i_3\rangle \langle i_2i_1\rangle
\langle i_1r_4\rangle}\cr
& + {\langle r_1 | \P_{i_3i_4} \P_{i_3i_4r_1r_2} \P_{r_3r_4}| i_1]^3\over
p^2_{r_3 r_4 i_1} p^2_{i_3 i_4 r_1} p^2_{i_3 i_4 r_1 r_2}
\, [r_3r_4] [r_4i_1]\, \langle i_3i_4\rangle \langle i_4 r_1\rangle
\,\langle i_2 | \P_{r_4i_1}|r_3]\, \langle i_3|\P_{i_4r_1}|r_2]
\langle i_2 | \P_{i_3i_4r_1} | r_2]}\cr
\noalign{\vskip4pt}
& + \;\hbox{ symmetry flip}.
\label{finalfour}\end{align}

 This is equivalent to the gauge theory expression \cite{RMV1, BCF, H1}. 
 \vskip15pt

For $(5,3)$, the four possible
step diagrams are drawn in  Figure 5.

a) The first diagram in Figure 5 corresponds to a residue evaluated on the poles
$f_{11} = f_{21}= f_{31} = 0$. This diagram has $p=0$,
with $n_0 = n =3$, $m_1 =4$ and $\bS_1 =\{r_3,i_1,i_2,i_3,i_4\}$.
From the general formula (\ref{genform}) the contribution to $\M_{5,3}$
yields the first term in (\ref{fivethree}).

b) The second diagram in Figure 5 corresponds to a residue evaluated at the poles
$f_{11} = f_{32}= f_{42} = 0$. This diagram has $p=1$,
with $n_0 = n =3$, $m_1 = 2$, $n_1=2$, $m_2=4$, and $\bS_1 =\{r_3,i_1,i_2\}$,
$\bS_2 =\{i_5,r_1\}$, $\bT_1 = \{r_2,r'_3, i'_2\}$,
where $\P_{\bT_1} = -\P_{\bT'_1}$, with $\bT'_1 = \{i_3,i_4,i_5,r_1\}$,
and contributes the second term in (\ref{fivethree}).

\setlength{\unitlength}{1.5mm}
\centerline{\hskip8truemm
\begin{picture}(16,24)
\linethickness{0.075mm}
\multiput(0,0)(4,0){4}{\line(0,1){20}}
\multiput(0,0)(0,4){6}{\line(1,0){12}}
\Red
\linethickness{0.3mm}
\multiput(2,6)(4,0){2}{\line(0,1){12}}
\multiput(2,6)(0,4){4}{\line(1,0){4}}
\Black
\end{picture}
\begin{picture}(16,24)
\linethickness{0.075mm}
\multiput(0,0)(4,0){4}{\line(0,1){20}}
\multiput(0,0)(0,4){6}{\line(1,0){12}}
\Red
\linethickness{0.3mm}
\multiput(2,14)(4,0){2}{\line(0,1){4}}
\multiput(2,14)(0,4){2}{\line(1,0){4}}
\Blue
\multiput(6,2)(4,0){2}{\line(0,1){8}}
\multiput(6,2)(0,4){3}{\line(1,0){4}}
\Black
\end{picture}
\begin{picture}(16,24)
\linethickness{0.075mm}
\multiput(0,0)(4,0){4}{\line(0,1){20}}
\multiput(0,0)(0,4){6}{\line(1,0){12}}
\Red
\linethickness{0.3mm}
\multiput(2,10)(4,0){2}{\line(0,1){8}}
\multiput(2,10)(0,4){3}{\line(1,0){4}}
\Blue
\multiput(6,2)(4,0){2}{\line(0,1){4}}
\multiput(6,2)(0,4){2}{\line(1,0){4}}
\Black
\end{picture}
\begin{picture}(16,24)
\linethickness{0.075mm}
\multiput(0,0)(4,0){4}{\line(0,1){20}}
\multiput(0,0)(0,4){6}{\line(1,0){12}}
\Blue
\linethickness{0.3mm}
\multiput(6,2)(4,0){2}{\line(0,1){12}}
\multiput(6,2)(0,4){4}{\line(1,0){4}}
\Black
\end{picture}
}

	\vskip6pt
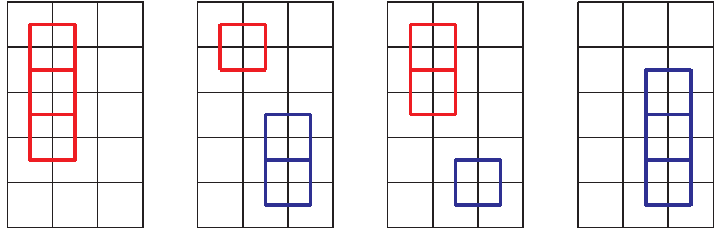
\centerline{Figure 5. \sl Step Diagrams for the $(5,3)$ Split Helicity Amplitude.}
\vskip6pt

The amplitude $\M_{5,3} (i_1,i_2,i_3,i_4,i_5,r_1,r_2,r_3)$ 
remains invariant under the transformation
$i_1\leftrightarrow i_5, i_2\leftrightarrow i_4, i_3\leftrightarrow i_3,
r_1\leftrightarrow r_3, r_2\leftrightarrow r_2$, and the 
 remaining two step diagrams are related
to the two we have computed above, by this transformation, the third to the second and the fourth to the first;
so altogether we find
\begin{align}
\M_{5,3}
&= {[i_5|\P_{r_1r_2} | r_3\rangle^3\over
p^2_{i_5r_1r_2} [i_5 r_1] [r_1r_2] \langle r_3 i_1\rangle\langle i_1i_2\rangle
\langle i_2i_3\rangle \langle i_3i_4\rangle \,
[ r_2 |\P_{i_5r_1} | i_4\rangle}\cr
&\qquad -{\langle r_1|\P_{i_3i_4i_5} \P_{i_1i_2}|r_3\rangle^3\over
p^2_{r_3i_1i_2} \, p^2_{i_3i_4i_5r_1} \, \langle i_3|\P_{i_4i_5r_1}| r_2]
\langle i_2|\P_{r_3i_1} | r_2]\,
\langle r_3i_1\rangle \langle i_1i_2\rangle\langle i_3i_4\rangle
\langle i_4 i_5\rangle \langle i_5r_1\rangle}\cr
\noalign{\vskip4pt}
&\qquad + \hbox{symmetry flip}.\label{fivethree}\end{align}
This is equivalent to the gauge theory expression
\cite{K1}-\cite{Luo}, \cite{Britto}. 

\section{\bf Non-Split Helicity Tree Amplitudes}
\label{nonsplit}

The methods developed here for calculating split helicity amplitudes can be extended to the non-split helicity case. It is not so obvious how to choose the analogue of contiguous constraints for non-split helicity amplitudes but we can infer an appropriate choice by using  a procedure for deriving the non-split integrand from the split integrand. The integrand function $\hat F^{NS}_{mn}(c)$ for the tree amplitude $(i_1, \ldots, i_a,r_b,r_{b-1}\ldots r_1, i_m,\ldots, i_{a+2}, i_{a+1}, r_{b+1},\ldots, r_n)$, where, as usual, the indices $i$ indicate positive helicities and the indices $r$ negative helicities, is related to the integrand function $\hat F_{mn}(c)$ for the split helicity case by 
\begin{align}
\hat F^{NS}_{mn}(c) 
&= {(\rho_{i_a}-\rho_{i_{a+1}})(\rho_{r_b}-\rho_{r_{b+1}})\over(\rho_{i_a}-\rho_{r_b})(\rho_{i_{a+1}}-\rho_{r_{b+1}})}\hat F_{mn}(c)\cr
&= {f_{ab}\over c_{a,b+1}c_{a+1,b}}\hat F_{mn}(c)\label{fns},
\end{align}
where $\rho_k$ denote the twistor string variables as described in \cite{DG3}. By repeating this procedure a number of times any non-split helicity amplitude can be obtained from a split helicity amplitude. The procedure (\ref{fns}) does not introduce more poles into $\hat F^{NS}_{mn}(c)$, beyond the those present in $\hat F_{mn}(c)$ as a result of the denominator $f_{11}f_{m-1,n-1}$ of (\ref{Fhat}), for most values of $a,b$ unless $m$ or $n$ is small. For example if $a=1, b=1$, so that we are discussing an amplitude with the helicity structure $(+, -, +, \ldots, +, -,\ldots -)$, we see from  (\ref{fns})  that the denominator of $\hat F^{NS}_{mn}(c)$ is just $f_{m-1,n-1}$, so that methods similar to those described in section \ref{RecRel} can be used.
\vss
\section*{Acknowledgements}
We are grateful to Nima Arkani-Hamed, Jacob Bourjaily, Freddy Cachazo, 
Dhritiman Nandan, 
Mark Spradlin, Jaroslav Trnka, 
Anastasia Volovich and Cong-kao Wen for discussions.
LD thanks the Institute for Advanced Study at Princeton for its hospitality. 
LD was partially supported by the U.S. Department of Energy, 
Grant No. DE-FG02-06ER-4141801, Task A.

\appendix

\section{Evaluations of Jacobians}
\label{Jacobians}

\subsection{\sl Residue Evaluation: the Jacobian $\displaystyle \left|{\partial  f\over\partial\hat\beta}\right|$
}
\label{ResEv}

We  consider the calculation of the Jacobian
\be{\partial  f\over\partial\hat\beta}\equiv{\partial 
( f_{11}, f_{12},\ldots, f_{1,n-2})\over\partial(\hat\beta_{22},
\hat\beta_{23},\ldots,\hat\beta_{2,n-1})}, \qquad \hbox{where} \qquad
 f_{1b}= c_{1b} c_{2,b+1}- c_{2b} c_{1,b+1}.\ee
From (\ref{cbb}),
\begin{align}
 c_{1b}&=a_{i_1r_b}+[i_2,i_3]\left(\langle r_{b-2},r_{b-1}\rangle
\hat\beta_{2,b-1}-\langle r_{b-1},r_{b+1}\rangle\hat\beta_{2,b}
+\langle r_{b+1},r_{b+2}\rangle\hat\beta_{2,b+1}\right)\cr
 c_{2b}&=a_{i_2r_b}-[i_1,i_3]\left(\langle r_{b-2},r_{b-1}\rangle\hat
\beta_{2,b-1}-\langle r_{b-1},r_{b+1}\rangle\hat\beta_{2,b}+\langle r_{b+1},
r_{b+2}\rangle\hat\beta_{2,b+1}\right)+\ldots,\nonumber\end{align}
where we have omitted terms in the last line that do not involve 
$\hat\beta_{2b'}$. 
Then, writing $K_b=[i_1,i_3] c_{1,b}+[i_2,i_3] c_{2,b}$,
the matrix  $\displaystyle{\partial  f\over\partial\hat\beta}$ is given by an 
$(n-2)\times (n-2)$ matrix beginning
\be\left(\begin{matrix}
\langle \varpi_{12},r_3\rangle& \langle \varpi_{23},r_1\rangle& K_4
\langle r_1,r_2\rangle&0&\ldots\cr
-K_1\langle r_3,r_4\rangle& \langle \varpi_{23},r_4\rangle&  
\langle \varpi_{34},r_2\rangle& K_5\langle r_2,r_3\rangle&\ldots\cr 
0& -K_2\langle r_4,r_5\rangle& \langle \varpi_{34},r_5\rangle&\langle 
\varpi_{45},r_3\rangle&\ldots\cr
0 &0 & -K_3\langle r_5,r_6\rangle& \langle \varpi_{45},r_6\rangle&\ldots\cr
\vdots&\vdots&\vdots&\vdots&\end{matrix}\right)\nonumber\ee
and ending
\be\left(\begin{matrix}
&\vdots&\vdots&\vdots&\vdots\cr
\ldots& \langle \varpi_{n-5,n-4},r_{n-3}\rangle& \langle\varpi_{n-4,n-3},
r_{n-4}\rangle& K_{n-2}\langle r_{n-5},r_{n-4}\rangle&0\cr
\ldots& K_{n-5}\langle r_{n-2},r_{n-3}\rangle& \langle\varpi_{n-4,n-3},
r_{n-2}\rangle&  \langle \varpi_{n-3,n-2},r_{n-4}\rangle
& K_{n-1}\langle r_{n-4},r_{n-3}\rangle\cr
\ldots&0 &  K_{n-4}\langle r_{n-1},r_{n-2}\rangle& \langle \varpi_{n-3,n-2},
r_{n-1}\rangle&  \langle \varpi_{n-2,n-1},r_{n-3}\rangle\cr
\ldots&0 &0 & K_{n-3}\langle r_{n},r_{n-1}\rangle& \langle \varpi_{n-2,n-1},
r_n\rangle
\end{matrix}\right),\nonumber\ee
with $\varpi_{ab}=K_a\pi_{r_a}+K_{a+1}\pi_{r_{a+1}}+\ldots+K_b\pi_{r_b}$.
If the corresponding Jacobian determinant is denoted by $J^f_n$, this gives the recurrence relationship

\begin{align}
J^f_n=\langle \varpi_{n-2,n-1},r_n&\rangle J^f_{n-1}-K_{n-3}\langle r_{n},
r_{n-1}\rangle\langle \varpi_{n-2,n-1},r_{n-3}\rangle J^f_{n-2}\cr
&+
K_{n-1} K_{n-3} K_{n-4}\langle r_{n},r_{n-1}\rangle\langle r_{n-1},
r_{n-2}\rangle\langle r_{n-4},r_{n-3}\rangle J^f_{n-3},
\label{recurx}\end{align}
which will determine $J^f_n$ subject to
\begin{align}
J^f_3&=\langle\varpi_{12},r_3\rangle, \qquad
J^f_4=K_2\langle r_2,r_3\rangle\langle\varpi_{13},r_4\rangle,\qquad
J^f_5=K_2K_3\langle r_2,r_3\rangle\langle r_3,r_4\rangle\langle\varpi_{14},
r_5\rangle.\nonumber\end{align}
We now show by induction that
\be J^f_n=\langle\varpi_{1,n-1},r_n\rangle \prod_{b=2}^{n-2}K_b\langle r_b,
r_{b+1}\rangle.\label{A}\ee

If (\ref{A}) holds, the right hand side of the recurrence relation 
(\ref{recurx}) is
\begin{align}
\left(\langle \varpi_{n-2,n-1},r_n\rangle \langle\varpi_{1,n-2},r_{n-1}
\rangle \right.\langle r_{n-3},&r_{n-2}\rangle-\langle r_{n},r_{n-1}
\rangle\langle \varpi_{n-2,n-1},r_{n-3}\rangle
\langle\varpi_{1,n-3},r_{n-2}\rangle\cr
+ K_{n-1} \langle r_{n},r_{n-1}\rangle\langle r_{n-1},r_{n-2}\rangle 
& \left.\langle\varpi_{1,n-4},r_{n-3}\rangle \right)K_{n-3} K_{n-4}
\langle r_{n-4},r_{n-3}\rangle\prod_{b=2}^{n-5}K_b\langle r_b,r_{b+1}\rangle.
\nonumber\\\label{bb}\end{align}
Now part of (\ref{bb}) is given by 
\begin{align}
\left(\langle \varpi_{n-2,n-1},r_n\rangle \langle\varpi_{1,n-4},r_{n-1}
\rangle \right.
\langle r_{n-3},&r_{n-2}\rangle-\langle r_{n},r_{n-1}\rangle\langle 
\varpi_{n-2,n-1},r_{n-3}\rangle \langle\varpi_{1,n-4},r_{n-2}\rangle
\cr
&+K_{n-1} \langle r_{n},r_{n-1}\rangle\langle r_{n-1},r_{n-2}\rangle  \left.
\langle\varpi_{1,n-4},r_{n-3}\rangle \right)\cr
&\hskip-40truemm=K_{n-2}\langle r_{n-3},r_{n-2}\rangle(\langle r_{n-2},
r_n\rangle \langle\varpi_{1,n-4},r_{n-1}\rangle+
\langle r_{n},r_{n-1}\rangle \langle\varpi_{1,n-4},r_{n-2}\rangle)\cr
&\hskip-40truemm=K_{n-2}\langle r_{n-3},r_{n-2}\rangle\langle r_{n-2},
r_{n-1}\rangle \langle\varpi_{1,n-4},r_n\rangle.\cr
\nonumber\end{align}
The remaining terms in (\ref{bb}) are
\begin{align}
\langle \varpi_{n-2,n-1},r_n\rangle& \langle\varpi_{n-3,n-2},r_{n-1}\rangle
\langle r_{n-3},r_{n-2}\rangle-\langle r_{n},r_{n-1}\rangle\langle 
\varpi_{n-2,n-1},r_{n-3}\rangle
K_{n-3}\langle r_{n-3},r_{n-2}\rangle \cr
&=K_{n-2} \langle\varpi_{n-2,n-1},r_n\rangle \langle r_{n-2},r_{n-1}\rangle 
\langle r_{n-3},r_{n-2}\rangle\cr
&\hskip20truemm+\langle \varpi_{n-2,n-1},r_{n-1}\rangle \langle r_{n-3},r_n\rangle
K_{n-3}\langle r_{n-3},r_{n-2}\rangle \cr
&=K_{n-2} \langle\varpi_{n-3,n-1},r_n\rangle \langle r_{n-2},r_{n-1}\rangle 
\langle r_{n-3},r_{n-2}\rangle
\nonumber\end{align}
so that (\ref{bb}) does indeed equal (\ref{A}).

In order to compare the form of this Jacobian with 
terms coming from the residues in section 3, let 
$ f_{1b}=0$, $1\leq b\leq n-2$, where
\be{ c_{1b}\over \langle i_1,r_n\rangle}={ c_{2b}\over \langle i_2,r_n
\rangle}={c^{12}_{bn}\over \langle i_1,i_2\rangle},\qquad
K_b={\langle r_n,i_1\rangle[i_1,i_3]+\langle r_n,i_2\rangle[i_2,i_3]\over
\langle r_n,i_2\rangle}  c_{2b}
={\langle r_n|P_{i_1i_2}|i_3]\over\langle r_n,i_2\rangle}  c_{2b},
\nonumber\ee
$$\varpi_{1,n-1}=\sum_{b=1}^{n-1}K_b\pi_{r_b}={\langle r_n|P_{i_1i_2}|i_3]
\over\langle r_n,i_2\rangle} (\pi_{i_2}-c_{2n}\pi_{r_n}).$$
So
$$\langle\varpi_{1,n-1},r_n\rangle=-\langle r_n|P_{i_1i_2}|i_3],$$
and
\begin{align}
\left|{\partial  f\over\partial\hat\beta}\right |=J^f_n&=-{\langle r_n|P_{i_1i_2}|i_3]^{n-2}\over\langle r_n,i_2\rangle^{n-3}}  
\prod_{b=2}^{n-2}c_{2b}\langle r_b,r_{b+1}\rangle\cr
&= {\langle r_n|P_{i_1i_2}|i_3]^{n-2}\langle i_1,i_2\rangle\langle r_n,i_1
\rangle\over\langle r_n,i_2\rangle^{n-1}}
\prod_{b=2}^{n-2}\langle r_b,r_{b+1}\rangle 
\left[{1\over c_{11} f_{1,n-1}}\prod_{b=1}^{n-1}c_{2b}\right].
\label{simform}\end{align}

\subsection{\sl Variable Change: the Jacobian
$\displaystyle\left|{\partial \hat\beta'\over\partial\hat\beta}\right |$}
\label{VarCh}

Had we not made the change of variables in the beginning of this section,
it would have been difficult to compute a general form for the 
Jacobian. However, we are not quite finished. 
If we use the conditions $ f_{1b}=0$, $1\leq b\leq n-2$, 
to eliminate $\pi_{i_1}, \bpi_{i_1},  c_{1b}, 1\leq b\leq n$, 
to define a new system as in section \ref{GenSol}(a), consisting of
\be\pi_{i_2},\;\bpi_{i_2}'={1\over \langle i_2,r_n\rangle}\P_{\bS}\pi_{r_n}; \quad
\pi_{i_a},\;\bpi_{i_a}, \,3\leq a\leq m;\ee
\be  \pi_{r_b},\;\bpi_{r_b}, \,
1\leq b\leq n-1;\;\quad
\pi_{r_n},\,\bpi_{r_n}'={1\over \langle r_n,i_2\rangle}\P_{\bS}\pi_{i_2}; \ee
where $\bS=\{i_1,i_2,r_n\}$, and $ c_{ab}$, $2\leq a\leq m, 1\leq b\leq n$, 
and then use this system to define a set of contiguous $\beta$'s, 
$\hat\beta'_{ab}$, 
$3\leq a\leq m-1, 2\leq b\leq n-1$,  these will not be the same as the 
corresponding $\hat\beta_{ab}$.  So we need to calculate the corresponding 
Jacobian. 
[Note that in section 4.1, we used the constraints to eliminate $\pi_{i_2}, 
\bpi_{i_2}$ and redefined $\bpi_{i_1}'$, but here we must rephrase this 
because the remaining constraints involve $ c_{2b}, 1\leq b\leq n$, 
and not $ c_{1b}$.]

To calculate the appropriate Jacobian, let 
\be B_{ab}= -\langle r_{b-2},r_{b-1}\rangle\hat\beta_{a,b-1}+\langle r_{b-1},
r_{b+1}\rangle\hat\beta_{ab}-\langle r_{b+1},r_{b+2}\rangle\hat\beta_{a,b+1},
\;3\leq b\leq n-2, \nonumber\ee
$$B_{a1}= -\langle r_{2},r_{3}\rangle\hat\beta_{a2},\quad
B_{a2}=\langle r_{1},r_{3}\rangle\hat\beta_{a2}-\langle r_{3},r_{4}
\rangle\hat\beta_{a3},$$
$$B_{a,n-1}= -\langle r_{n-3},r_{n-2}\rangle\hat\beta_{a,n-2}+\langle r_{n-2},
r_{n}\rangle\hat\beta_{a,n-1},\quad
B_{an}= -\langle r_{n-2},r_{n-1}\rangle\hat\beta_{a,n-1},$$
and similarly for $B_{ab}'$ in terms of $\hat\beta_{ab}'.$ Then
$$ c_{1b}=a_{1b}-[i_2,i_3]B_{2b}, \quad  c_{2b}=a_{2b}+[i_1,i_3]B_{2b}
-[i_3,i_4]B_{3b}, $$
$$  c_{3b}=a_{3b}-[i_1,i_2]B_{2b}+[i_2,i_4]B_{3b}
-[i_4,i_5]B_{4b}.$$
So, from $ c_{1b}\langle i_2,r_n\rangle= c_{2b} \langle i_1,r_n\rangle$,
$$\langle r_n,i_1\rangle [i_3,i_4]B_{3b}=\langle r_n|P_{i_1i_2}|i_3]B_{2b}
+\langle r_n,i_1\rangle a_{2b}-\langle r_n,i_2\rangle a_{1b}$$
$$ c_{ 2 b}=a_{ 2 b}'-[i_3,i_4]B_{3b}', \quad  c_{3b}=a_{3b}'
+[i_1',i_4]B_{3b}'-[i_4,i_5]B_{4b}',$$
$${\langle r_n,i_1\rangle\over\langle r_n,i_2\rangle}
[i_3,i_4]B_{3b}'=[i_2,i_3]B_{2b}+{\langle r_n,i_1\rangle\over\langle i
r_n,i_2\rangle} a_{1b}'-a_{1b}$$
$$\langle r_n|P_{i_1i_2}|i_3]B_{3b}'=[i_2,i_3]\langle r_n,i_{ 2}\rangle B_{3b}
+\ldots$$
from which it follows that
$$\beta'_{3b}={[i_2,i_3]\langle r_n,i_1\rangle\over\langle r_n|P_{i_1i_2}|i_3]}
\beta_{3b}+\varphi_{3b},\qquad\beta'_{ab}
=\beta_{ab}+\varphi_{ab},\quad 4\leq a\leq m-1,\;2\leq b\leq n-1,$$
where $\varphi_{ab}$ depends on the momenta and the $\beta_{a'b}$, $a'<a$, 
and, hence,
\be\left|{\partial \hat\beta'\over\partial\hat\beta}\right|=\left[{[i_2,i_3]
\langle r_n,i_{ 2}\rangle\over \langle r_n|P_{i_1i_2}|i_3]}\right]^{n-2}.
\label{jac3}\ee

%%%%%%%%%%%%%%%%%%%%%%%%%%%%%%%%ENDPAPER%%%%%%%%%%%
\vfill\eject

\singlespacing

%%% References %%%

\providecommand{\bysame}{\leavevmode\hbox to3em{\hrulefill}\thinspace}
\providecommand{\MR}{\relax\ifhmode\unskip\space\fi MR }
\providecommand{\MRhref}[2]{%
  \href{http://www.ams.org/mathscinet-getitem?mr=#1}{#2}
}
\providecommand{\href}[2]{#2}

\end{document}